%% file: main.tex
\documentclass[conference,table]{IEEEtran}

\usepackage[hyphens]{url}
\usepackage{booktabs}
\let\labelindent\relax
\usepackage[hidelinks]{hyperref}

\usepackage{algorithm}
\usepackage{algorithmic}
\usepackage{amsmath}
\usepackage{amsthm}
\usepackage{array}
\usepackage{dsfont}
\usepackage{enumitem}
\usepackage{float}
\usepackage{makecell}
\usepackage{multirow}
\usepackage[available,functional,reproduced]{ndssbadges}
\usepackage{svg}
\usepackage{xcolor}
\usepackage{xspace}
\usepackage{vcell}
\usepackage{tabularray}
\usepackage{adjustbox} 

\DeclareMathOperator*{\argmax}{arg\,max}
\DeclareMathOperator*{\argmin}{arg\,min}
\DeclareMathOperator{\rank}{\mathbf{rank}}

\newtheorem{theorem}{Theorem}

\newcommand{\name}{RAIFLE\xspace}
\newcommand{\repo}{{\urlstyle{tt}\url{https://github.com/dzungvpham/raifle}\xspace}}
\def\code#1{\texttt{#1}}

\begin{document}

\title{\name: Reconstruction Attacks on Interaction-based Federated Learning with Adversarial Data Manipulation}

\author{
\IEEEauthorblockN{Dzung Pham, Shreyas Kulkarni, Amir Houmansadr}
\IEEEauthorblockA{University of Massachusetts Amherst\\
\{dzungpham, svkulkarni, amir\}@cs.umass.edu}
}

\IEEEoverridecommandlockouts
\makeatletter\def\@IEEEpubidpullup{6.5\baselineskip}\makeatother
\IEEEpubid{\parbox{\columnwidth}{
    Network and Distributed System Security (NDSS) Symposium 2025\\
    24-28 February 2025, San Diego, CA, USA\\
    ISBN 979-8-9894372-8-3\\
    https://dx.doi.org/10.14722/ndss.2025.240363\\
    www.ndss-symposium.org
}
\hspace{\columnsep}\makebox[\columnwidth]{}}

\maketitle

\begin{abstract}
Federated learning has emerged as a promising privacy-preserving solution for machine learning domains that rely on \emph{user interactions}, particularly recommender systems and online learning to rank.
While there has been substantial research on the privacy of traditional federated learning, little attention has been paid to the privacy properties of these interaction-based settings.
In this work, we show that users face an elevated risk of having their private interactions reconstructed by the central server when the server can control the training features of the items that users interact with. 
We introduce \name, a novel optimization-based attack framework where the server actively manipulates the features of the items presented to users to increase the success rate of reconstruction.
Our experiments with federated recommendation and online learning-to-rank scenarios demonstrate that \name is significantly more powerful than existing reconstruction attacks like gradient inversion, achieving high performance consistently in most settings.
We discuss the pros and cons of several possible countermeasures to defend against \name in the context of interaction-based federated learning.
Our code is open-sourced at \repo.
\end{abstract}

\input{intro}

\input{background}

\input{raifle}

\input{impl_details}

\input{evaluation_setup}

\input{evaluation_results}

\input{countermeasures}

\input{discussion}

\input{conclusion}

\section*{Acknowledgment}
We would like to thank Professor Negin Rahimi for her helpful comments on learning-to-rank and information retrieval. The project was supported in part by NSF grant 2131910.

\bibliographystyle{IEEEtranS}
\bibliography{references}

\input{appendix}

\end{document}

%% file: intro.tex
\section{Introduction}
Federated learning (FL)~\cite{mcmahan2017, kairouz2021} is an emerging approach to building privacy-preserving recommender systems (RS)~\cite{yang2020fedrec, sun2022fedrecsurvey} and online learning to rank (OLTR) solutions~\cite{kharitonov2019, wang2021fpdgd}.
In such systems, users interact with server-prepared ``items'' (e.g., news articles, media, and products) via clicks, ratings, and other types of interactions, then train their FL model using these private interactions.
As a result, users can benefit from a better item ranking/recommendation experience without having to share potentially sensitive interaction data with service providers.
We broadly define this variation on FL as \emph{interaction-based FL} (IFL).
Unlike traditional FL, users in IFL do not inherently own any data other than their interactions with the server-controlled items.
As the two most prominent IFL instances, RS and OLTR have many far-reaching applications such as web search (e.g., Google), online advertising (e.g., Facebook ads), and e-commerce (e.g., Amazon), which often rely on vast amounts of private and sensitive user data.
Consequently, the use of IFL for RS/OLTR has gained popularity in both academia and industry, with a wide variety of algorithms proposed such as federated collaborative filtering~\cite{ammad2019, perifanis2022fncf}, federated graph neural net for recommendation~\cite{wu2022fedgnn}, reinforcement learning-based federated OLTR~\cite{kharitonov2019}, etc.
Their potential to leverage data from diverse sources while maintaining user privacy and data ownership makes them a compelling approach to enhancing recommendation/ranking accuracy and addressing data silo challenges.

\begin{figure}
    \centering
    \includegraphics[width=\columnwidth]{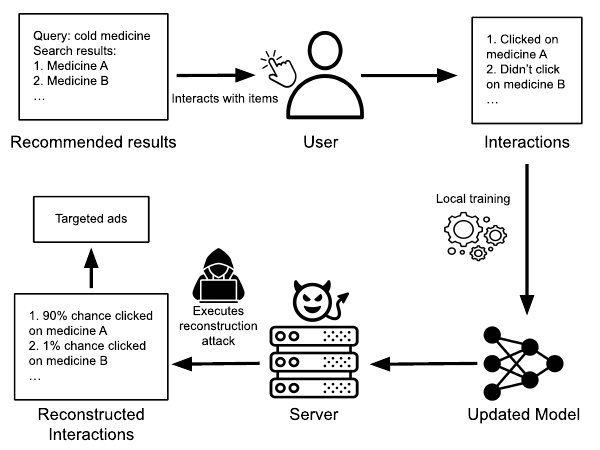}
    \caption{Example of an interaction reconstruction attack in federated recommendation/learning-to-rank. A malicious server may infer user interactions from the FL updates to execute targeted advertising.}
    \label{fig:rs}
\end{figure}

IFL's departure from standard FL thus poses a significant privacy risk that is currently not well understood.
Previous work on federated RS/OLTR~\cite{sun2022fedrecsurvey, yuan2023imia, wang2021fpdgd} has often overlooked the server's knowledge and control over the items presented to users.
This unique aspect of IFL is also not considered in existing FL privacy attacks like gradient inversion~\cite{huang2021evalgradinv, zhang2022surveygradinv, ovi2023gradinv}, since in traditional FL, the server usually cannot exert a great deal of influence on users' data or behaviors.
We argue that the IFL server should not require access to user interactions for two key reasons.
Firstly, the server does not always need user interaction data to provide essential services.
For example, web search engine providers do not need to know precisely which links users click on once they have delivered the search results to the users.
Secondly, the server can either accidentally or intentionally leak private user interactions and preferences, resulting in serious consequences such as compromising user anonymity~\cite{narayanan2008netflix}, exposing sensitive interests~\cite{weinsberg2012blurme}, and enabling targeted advertising~\cite{toubiana2010adnostic}, ultimately leading to the loss of user trust (Figure \ref{fig:rs}).
Addressing these privacy challenges is crucial to ensure the responsible and secure implementation of FL for RS and OLTR.

In this work, we aim to study the risk of reconstructing user interaction data in IFL.
We present \textbf{\name},\footnote{Pronounced like ``rifle''} a general \textbf{r}econstruction \textbf{a}ttack for \textbf{i}nteraction-based \textbf{f}ederated \textbf{le}arning systems.
Similar to the gradient inversion attacks in traditional FL~\cite{huang2021evalgradinv, zhang2022surveygradinv, ovi2023gradinv}, \name aims to find the most likely interactions by minimizing the ``distance'' between the simulated local update created with candidate interactions and the actual received local update.
However, unlike gradient inversion, \name exploits the IFL server's knowledge and control over the items to execute \emph{Adversarial Data Manipulation} (ADM), a novel attack vector where the server modifies the training data features associated with the items to produce adversarial behaviors in the local updates.
We design two ADM techniques: the \emph{fingerprint} method where the IFL server selectively chooses which training features to keep or ``zero out'' to control the local gradients and the \emph{noise injection} method where the training features are replaced with random noise.
We empirically show that this active adversarial attack can outperform vanilla gradient inversion as well as undermine existing privacy-enhancing mechanisms such as secure aggregation~\cite{bonawitz2017} and private information retrieval (PIR)~\cite{chor1998pir, kushilevitz1997pir}.
Our attack can also work even when the server does not have direct control over the training features, particularly when the users extract the features themselves.
To the best of our knowledge, \name is the first optimization-based active reconstruction attack applicable to not only federated RS/OLTR but also the more general IFL setting.

In light of these findings, we discuss the pros and cons of various potential countermeasures to alleviate the risks of our privacy attacks, including local differential privacy, secure aggregation, data validation, personalization, etc.
We hope our research can inform future endeavors to develop more secure and private federated RS/OLTR systems specifically and IFL in general.
To summarize, our main contributions are:
\begin{itemize}
    \item We identify the server's knowledge and control over the data items in interaction-based FL (IFL) scenarios like federated RS/OLTR as a privacy vulnerability that can facilitate stronger reconstruction attacks.
    \item We introduce \name, a general optimization-based reconstruction attack framework for IFL. \name utilizes Adversarial Data Manipulation (ADM), a novel attack vector unique to IFL where the IFL server actively manipulates the items to increase \name's attack performance.
    \item We evaluate \name in two representative IFL systems, namely federated RS and OLTR, and show that our attack has strong inference performance under most tested scenarios, even when the server can only indirectly influence the training features.
    \item We analyze various countermeasures to mitigate privacy leakage against \name and ADM for federated RS/OLTR and IFL in general.
\end{itemize}

%% file: background.tex
\section{Background and Related Work} \label{sec:background}

We provide a brief overview of privacy attacks and defenses in FL and a description of IFL.

\subsection{Attacks on FL} \label{sec:grad_inv}

\subsubsection{Passive Attacks}
Federated Learning (FL) is a decentralized approach to machine learning that allows each participant to collaboratively train a machine learning model without having to share their private data with a central server~\cite{mcmahan2017, kairouz2021}.
Despite the decentralization of data, research has shown that FL can be vulnerable to a class of attack called \textbf{gradient inversion}~\cite{huang2021evalgradinv, zhang2022surveygradinv, ovi2023gradinv}, which allows an honest-but-curious server to invert users' gradients to find an approximation of users' local data by solving an optimization problem of the following form:
\begin{equation} \label{eq:gradinversion}
    \operatorname*{argmin}_{\mathcal{X}'} \left[\textbf{dist}(\nabla^{\mathcal{X}'}_{\theta}, \nabla^{\mathcal{X}}_{\theta}) + \rho(\mathcal{X}')\right]
\end{equation}
where $\mathcal{X}'$ is the server's approximation of the user's data, $\mathcal{X}$ is the user's actual data, $\theta$ represents the model parameters, $\nabla^{\mathcal{X}'}_{\theta}$ and $\nabla^{\mathcal{X}}_{\theta}$ are the gradients w.r.t.\  $\theta$ when trained on $\mathcal{X}'$ and $\mathcal{X}$ (respectively),
$\textbf{dist}$ is a measure of distance between the gradients, and $\rho$ is the prior/regularizer placed on $\mathcal{X}'$.
Essentially, the server tries to find an approximation $\mathcal{X}'$ of the local data $\mathcal{X}$ that would lead to the simulated gradients $\nabla^{\mathcal{X}'}_{\theta}$ closest to the actual received gradients $\nabla^{\mathcal{X}}_{\theta}$.
Hence, the attack is also called ``gradient matching''.

The majority of research in gradient inversion has so far been focused on deep neural networks for computer vision and natural language processing tasks~\cite{ovi2023gradinv}, which differ from RS and OLTR in the following ways: (a) user interactions in RS and OLTR research are typically structured discrete data (e.g., clicks vs no clicks), unlike images and texts; (b) RS/OLTR models can have much fewer parameters and are also not restricted to neural nets.
These differences can render existing gradient inversion attacks not immediately applicable to the RS/OLTR context.

\subsubsection{Active Attacks}
Unlike gradient inversion, which only passively observes the local updates without changing any aspects of the FL protocol, active FL attacks involve a malicious server that deliberately modifies parts of the FL training process to further enable user data reconstruction.
Most notable is \textbf{model manipulation} attacks~\cite{fowl2021robbing, pasquini2022, boenisch2023reconstruct}, in which the server directly manipulates the FL model's architecture or weights in such a way that the user data is more easily leaked through the FL updates.
This class of attack has a much higher reconstruction quality than pure gradient inversion and can even be applied to FL with secure aggregation defenses.
Another type of active attack is Sybil-based attacks, where the server introduces fake FL participants that are completely under the server's control~\cite{boenisch2023sybil}.
By choosing only one real user and setting the FL updates from all other fake users to 0, the server can easily isolate the real user's local updates, thus bypassing both secure aggregation and differential privacy.

\subsubsection{Other Attacks} Deep neural networks (DNN) for computer vision tasks have been demonstrated to be vulnerable to \textbf{adversarial perturbations} in their image inputs, which can cause the models to diverge from expected behaviors~\cite{yuan2019adv, akhtar2018adv, akhtar2021adv, khamaiseh2022adv}.
An image's deep representations extracted from a DNN can in fact be manipulated via this attack to closely resemble those of any arbitrary image~\cite{sabour2016advrep}.
DNNs for natural language processing tasks are also susceptible to adversarial inputs, although the attack techniques are rather different from image-based attacks due to the discrete nature of the input space~\cite{zhang2020adv, wei2023jailbreak}.
While adversarial perturbation attacks are not commonly applied in traditional FL, it is particularly applicable to our data manipulation techniques for IFL given the server's control of the interaction items.
Another attack closely related to our problem of reconstructing user interactions is \textbf{membership inference attacks} (MIA) against ML models~\cite{shokri2017mia, nasr2019mia}, which aims to determine whether a record is part of the training data.

\subsection{Privacy Defenses for FL}

\subsubsection{Differential Privacy}
Considered the state-of-the-art privacy model and defense in statistics and machine learning, differential privacy (DP)~\cite{dwork2014, abadi2016} has been applied in numerous FL applications~\cite{mcmahan2018, ramaswamy2020, kairouz2021}.
Formally, a randomized mechanism $\mathcal{M}: \mathcal{X} \to \mathcal{Z}$ is $(\varepsilon, \delta)$-differentially private if for any pair of database $x, x' \in \mathcal{X}$ differing in \emph{at most} one record and for any $S \subseteq \mathcal{Z}$, we have:
\begin{equation} \label{dp_def}
    \mathds{P}[\mathcal{M}(x) \in S] \leq e^\varepsilon \mathds{P}[\mathcal{M}(x') \in S] + \delta
\end{equation}

where $\varepsilon \geq 0$ represents the privacy budget and $\delta \in [0, 1)$ represents the probability of privacy leakage. 
In FL, the randomized mechanism $\mathcal{M}$ is typically the training process that occurs on users' devices and the aggregation process.

The definition above describes the \emph{central} DP model, which in the context of FL means that users trust a third-party curator with the collection, aggregation, and privatization of their local updates before releasing them to the server.
In this paper, we rely on the stricter but more realistic \textbf{local differential privacy} model (LDP), where users do not trust any entity with their data~\cite{kasiviswanathan2008}. 
Formally, LDP requires equation $\ref{dp_def}$ to hold for \emph{any} pair of $x, x'$.
While LDP can effectively defend against gradient inversion attacks~\cite{zhang2022surveygradinv}, it often incurs a high utility cost~\cite{wei2020}.
Practical levels of $\varepsilon$ for LDP therefore tend to be ``in the hundreds''~\cite{bhowmick2019, google2023} in order to achieve a decent model utility.
Achieving a good balance between utility and privacy thus remains a challenging problem~\cite{cummings2024advancing}.

\subsubsection{Secure Aggregation}
Another line of privacy defense for FL is secure aggregation (SA)~\cite{bonawitz2017, bell2020}, which relies on cryptography techniques to securely aggregate (e.g., summing) the local updates before sharing them with the server.
With this mechanism, the server can only observe the aggregated result without knowing the individual contributions from any user. SA, therefore, aims to achieve two objectives: ``privacy by aggregation'' -- to make it difficult for the server to infer useful information about individuals from the aggregated updates -- and ``privacy by shuffling'' -- to hide the link between individuals and updates so that even if any information can be inferred, it cannot be traced back to any particular user~\cite{pasquini2022}.
Nevertheless, research has shown that vanilla SA for FL is susceptible to model manipulation attacks~\cite{fowl2021robbing, pasquini2022, boenisch2023reconstruct}, as the final aggregated model can be adversarially engineered to capture private local data without being affected by the aggregation.
To prevent such attacks with more theoretical guarantees, SA needs to be combined with user-applied noise to achieve distributed differential privacy~\cite{kairouz2021sa}, which requires less noise than pure LDP and thus better preserves FL utility.

\subsection{Interaction-based FL (IFL)} \label{sec:ifl}

\begin{figure}
    \centering
    \includegraphics[width=0.9\columnwidth]{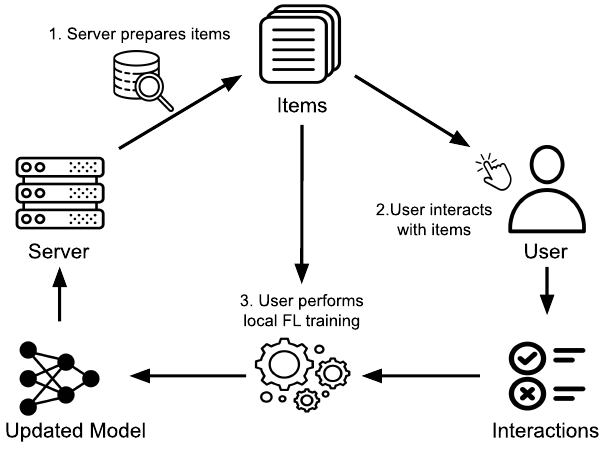}
    \caption{Diagram of Interaction-based Federated Learning (IFL). Users interact with server-prepared items and train the FL model using the items and their private interactions. Users may apply privacy defense techniques such as differential privacy before sending local updates to the server.}
    \label{fig:ifl}
\end{figure}

We broadly define interaction-based FL (IFL) as a variation of FL in which users \emph{interact} with items provided by the server and perform the local training using their interaction information (Figure \ref{fig:ifl}).
The nature of the items depends on the specific application (e.g., news articles for news recommendations or search results for web search).
Unlike traditional FL, IFL has two distinctive characteristics: (1) the server can influence how the items are (re)presented to the users, and (2) the primary data to be protected is the user interactions.
We describe below two major instances of IFL: federated recommendation systems and federated online learning to rank.

\subsubsection{Federated Recommender Systems}

Recommender systems (RS) is a technology that assists users with discovering relevant items like products, news, media, etc.~\cite{ricci2022}.
RS typically relies on user-item interactions, which unfortunately have been shown to be capable of revealing sensitive user attributes (e.g., age, gender)~\cite{weinsberg2012blurme} and even uniquely identifying users~\cite{narayanan2008netflix}.
To keep user interactions protected, recent research has looked to FL as the foundation to build more privacy-preserving recommendation services~\cite{yang2020fedrec, sun2022fedrecsurvey}.

One of the earliest and most influential RS algorithms is \textbf{collaborative filtering} (CF), which recommends items based on the behaviors and preferences of similar users~\cite{goldberg1992}.
A common approach to CF is \emph{matrix factorization}~\cite{canny2002}: essentially, the user-item interaction matrix is factored into two lower-dimensional latent factor matrices, one for the user embeddings and one for the item embeddings.
Building upon CF, \textbf{federated collaborative filtering} (FCF) formulates the item matrix as public parameters (i.e., known by all participants) and the user matrix as private parameters, which are privately updated by each user without being shared with anyone else~\cite{ammad2019}.
This idea has been incorporated in various subsequent works such as federated neural collaborative filtering (FNCF)~\cite{perifanis2022fncf} and federated graph neural net for recommendation (FedGNN)~\cite{wu2022fedgnn}.
However, keeping user embeddings private by itself does not automatically guarantee privacy for users since their interactions can still be inferred as demonstrated by previous research \cite{chai2021, yuan2023imia} (as well as our own work).

\subsubsection{Federated Online Learning to Rank} \label{sec:fedrs}

Learning to rank (LTR) is a machine learning task whose aim is to learn ranking models for information retrieval systems and has been applied to a variety of applications such as recommendation and web search~\cite{liu2009ltr}.
There are three major LTR approaches: (a) the \emph{pointwise} approach compares each item to a relevance label, (b) \emph{pairwise} compares pairs of items with different preferences, and (c) \emph{listwise} considers the entire ranked list as a unit to optimize the overall ranking.
In this paper, we focus on federated online learning to rank (FOLTR), which primarily learns from user interactions (i.e., implicit feedback), unlike traditional LTR which relies on labeled (i.e., explicit) relevance judgement~\cite{grotov2016oltr, kharitonov2019, wang2021fpdgd}.
Of particular interest is the \textbf{Federated Pairwise Differentiable Gradient Descent} (FPDGD) method, which proposes using the Pairwise Differentiable Gradient Descent (PDGD) algorithm~\cite{wang2021fpdgd, oosterhuis2018} in the FL setting.
The authors of FPDGD claim a higher ranking performance than the existing state-of-the-art FOLTR algorithm~\cite{kharitonov2019}, even with DP applied.
(We note, however, that FPDGD's evaluation with DP is not fully substantiated as it contains two issues: (a) the Laplace mechanism used only satisfies central DP, whereas the mechanism used in \cite{kharitonov2019} satisfies LDP, and (b) the authors clipped the $L_2$-norm of the model weights instead of $L_1$-norm as required by the Laplace mechanism.)

\subsubsection{Attacks on IFL}
To the best of our knowledge, only one paper has attempted to devise a privacy attack to steal user interactions in federated RS~\cite{yuan2023imia}.
However, their proposed Interaction Membership Inference Attack (IMIA) is a heuristic-based search that does not make use of the gradients with respect to the simulated interactions, resulting in poor inference performance.
Furthermore, the attack is limited to only federated RS and cannot be directly applied to other IFL scenarios such as OLTR.

%% file: raifle.tex
\section{Overview of \name} \label{sec:overview}

In this section, we describe the high-level approach of \name, our reconstruction attack on IFL.
Descriptions of useful notations used throughout the subsequent sections are provided in Table \ref{tab:notations}.

\begin{table}
\small
\centering
\caption{Notations}
\label{tab:notations}
\begin{tabular}{@{}c  p{7cm}@{}}
\toprule
Symbol & Description \\
\midrule
$m$ & Number of items \\
$n$ & Number of users \\
$d$ & Dimension of item features or representations \\
$p$ & Number of model parameters, assumed to be $\geq d$ \\
$\mathcal{X}$ & A matrix in $\mathds{R}^{m \times d}$ that represents training features (or embeddings for RS) of the items \\
$\hat{\mathcal{X}}$ & The user's updated item embeddings for RS \\
$\mathcal{I}$ & A vector in $\mathds{R}^{m}$ for the user's true interactions \\
$\mathcal{I}'$ & The server's reconstructed interactions \\
$\theta$ & A vector in $\mathds{R}^{p}$ for the global model parameters \\
$\hat{\theta}$ & The user's updated model parameters \\
$f$ & A function that represents the global model. Takes $\mathcal{X}$, $\theta$, $\mathcal{I}$. Returns the output of the model. \\
$g$ & The local FL algorithm. Takes $\mathcal{X}, \theta, \mathcal{I}$ (and user embedding in RS). Returns the updated model parameters (and the updated item embeddings for RS). \\
$\mathcal{L}_{atk}$ & Loss function used by our attack \\
$\nabla_{\mathcal{I'}} g$ & A $m \times p$ matrix for the gradients of $g$ w.r.t.\ $\mathcal{I'}$ \\
$\nabla_{\mathcal{I'}}^2 g$ & A third-order tensor for the Hessian of $g$ w.r.t.\ $\mathcal{I'}$ \\
$\nabla_{\mathcal{I'}} \mathcal{L}_{atk}$ & $m \times 1$ vector for the gradients of $\mathcal{L}_{atk}$ w.r.t.\ $\mathcal{I'}$ \\
$\nabla_{\mathcal{I'}}^2 \mathcal{L}_{atk}$ & $m \times m$ matrix for the Hessian of $\mathcal{L}_{atk}$ w.r.t.\ $\mathcal{I'}$ \\
\bottomrule
\end{tabular}
\end{table}

\subsection{Threat Model} \label{raifle:threat}

Our adversary is a FL server that not only knows which items are presented to which users but can also modify the representations of the items served to users, particularly the training features used in the local FL training.
This threat model is consistent with previous research that involves active and malicious FL servers~\cite{bonawitz2017, bell2020, pasquini2022}.
The server's knowledge of user-item impressions is realistic in most practical IFL systems like federated RS and OLTR.
While it is technically possible to hide the user-item mapping via Private Information Retrieval (PIR)~\cite{chor1998pir}, the prohibitively expensive computational costs prevent such a technique from being widely adopted at scale.
The ability to modify items is also justifiable for various business/operational reasons, such as the need to experiment with new/updated item features to improve service quality.
 
\subsection{Basic Reconstruction Framework} \label{sec:raifle_framework}

We first present a high-level approach to \name without any server-side manipulation from the Bayesian perspective.
Given a model $f$ with global parameters $\theta$, a user $u$, the items $\mathcal{X}$ presented to $u$, and $u$'s updated model parameters $\hat{\theta}$ learned via a learning algorithm $g$, we are interested in inferring $u$'s true interactions $\mathcal{I}$.
Probabilistically speaking, we want to find:
\begin{equation} \label{eq:raifle1}
    \argmax_{\mathcal{I}'} Pr(\mathcal{I}' \mid \mathcal{X}, \theta, \hat{\theta})
\end{equation}

Applying Bayes' theorem, we get:
\begin{equation} \label{eq:raifle2}
Pr(\mathcal{I}' \mid \mathcal{X}, \theta, \hat{\theta})
\propto Pr(\hat{\theta} \mid \mathcal{X}, \theta, \mathcal{I}') \cdot Pr(\mathcal{I}' \mid \mathcal{X}, \theta)
\end{equation}

Thus, we can approach the problem as a maximum a priori (MAP) estimation task, where we try to maximize the right-hand side (RHS) of equation \ref{eq:raifle2}.
$\hat{\theta}$ can be considered the ``data'', $\mathcal{I'}$ can be considered the ``parameters'', and the second term of the RHS can be considered the prior on $\mathcal{I'}$.
In our experiments, we use an uninformative (e.g., uniform) prior and focus on maximizing the first term of the RHS, i.e., the likelihood of $\hat{\theta}$ given $\mathcal{X}, \theta, \mathcal{I}'$ via optimization:
\begin{equation}
    \argmin_{\mathcal{I}'} \mathcal{L}_{atk}(\hat{\theta}, g(\mathcal{X}, \theta, \mathcal{I'}))
\end{equation}
\noindent where $\mathcal{L}_{atk}$ is a loss function that measures the ``distance'' (e.g., $L_2$) between the user's learned parameters $\hat{\theta}$ and the server's simulated parameters.

If the interaction values are discrete (e.g., clicked vs. no clicked, integer rating), one approach is to brute-force through all possible $\mathcal{I'}$ to find the one that would produce the closest simulated parameters to $\hat{\theta}$, but this is not computationally feasible when $|\mathcal{I'}|$ is large.
To enable a gradient-based attack, the server needs to make $g$ be (twice-)differentiable w.r.t.\ $\mathcal{I}'$.
Automatic differentiation (Autodiff)~\cite{baydin2017ad} can then be employed to calculate $\nabla_{\mathcal{I}'} \mathcal{L}_{atk}$.
Users can still use the original $g$ since \name takes place on the server only. If $g$ is a gradient-based learning algorithm, then our attack can also be considered ``gradient matching'' (eq. \ref{eq:gradinversion}).
Note that we do not require $g$ to be differentiable w.r.t.\ $\theta$ for \name to work as long as $g$ is differentiable w.r.t. $\mathcal{I'}$.
Furthermore, under specific circumstances, \name is \textbf{convex} and can have a \textbf{unique global optimum} as we formally prove below:

\begin{theorem}[Convexity of \name] \label{thm:convexity}
Assume that $g$ is twice-differentiable w.r.t.\ interactions $\mathcal{I'}$ and $\mathcal{L}_{atk}$ is the $L_2$ loss. If $\nabla_{\mathcal{I'}}^2 g = \mathbf{0}$, then \name is convex w.r.t.\ $\mathcal{I'}$.
\end{theorem}

\begin{proof}
    Consider the gradient of $\mathcal{L}_{atk}$ w.r.t.\ any $\mathcal{I'}$:
    \begin{equation} \label{eq:grad}
        \begin{aligned}
            \nabla_{\mathcal{I'}} \mathcal{L}_{atk} (\hat{\theta}, \theta')
            &= \nabla_{\mathcal{I'}} || \hat{\theta} - g(\mathcal{X}, \theta, \mathcal{I'}) ||^2_2 \\
            &= -2 \nabla_{\mathcal{I'}} g(\mathcal{X}, \theta, \mathcal{I'}) \cdot (\hat{\theta} - g(\mathcal{X}, \theta, \mathcal{I'})) \\
        \end{aligned}
    \end{equation}

    Thus, the Hessian of $\mathcal{L}_{atk}$ w.r.t.\ $\mathcal{I'}$ is:
    \begin{equation} \label{eq:hessian}
        \begin{aligned}
            &\nabla_{\mathcal{I'}}^2 \mathcal{L}_{atk} \\
            = &-2 \nabla_{\mathcal{I'}}^2 g(\mathcal{X}, \theta, \mathcal{I'}) \cdot \hat{\theta} + 2 \nabla_{\mathcal{I'}} g(\mathcal{X}, \theta, \mathcal{I'}) \cdot \nabla_{\mathcal{I'}}^T g(\mathcal{X}, \theta, \mathcal{I'}) \\
            = &2 \nabla_{\mathcal{I'}} g(\mathcal{X}, \theta, \mathcal{I'}) \cdot \nabla_{\mathcal{I'}}^T g(\mathcal{X}, \theta, \mathcal{I'}) \\
        \end{aligned}
    \end{equation}

    Clearly, $\nabla_{\mathcal{I'}}^2 \mathcal{L}_{atk} \succeq \mathbf{0}\ \forall \mathcal{I'} \in \mathds{R}^{m}$ (i.e., positive semi-definite). We also know that $\mathcal{I'} \in \mathds{R}^{m}$ is a convex set. Therefore, by the second-order convexity condition~\cite{boyd2004}, \name is convex w.r.t.\ $\mathcal{I'}$.
\end{proof}

From Eq. \ref{eq:hessian}, we can see that (under the same conditions above) \name is strictly convex if and only if $\rank(\nabla_{\mathcal{I'}} g) = m$, and consequently, our attack can arrive at a unique solution that minimizes $\mathcal{L}_{atk}$.
While making the learning algorithm $g$ twice-differentiable w.r.t. $\mathcal{I'}$ is easy, requiring $\nabla_{\mathcal{I'}}^2 g = \mathbf{0}$ is not always possible if the interactions ``interact'' with one another in $g$ (see Section \ref{raifle:diff}).
To enable $\rank(\nabla_{\mathcal{I'}} g) = m$, the server can potentially manipulate the data (and the learning algorithm $g$) to keep the number of items $m$ smaller than the number of training features $d$ or model parameters $p$ and remove any collinearity in the training features.
It should be noted that even when the attack can provably converge to a unique optimal solution, there is no formal guarantee that the solution is the actual user interactions since we do not have any guarantee about the user's local updates $\hat{\theta}$.

\subsection{Adversarial Data Manipulation} \label{sec:manipulation_approach}

We further enhance the basic RAIFLE framework with a novel attack vector called Adversarial Data Manipulation (ADM), in which the FL server actively manipulates the item features presented to users to adversarially improve the reconstruction success rate.
This technique is unique to IFL and does not apply to traditional FL since users in IFL interact with data prepared by the server.
We present two specific ADM methods called \emph{fingerprinting} and \emph{noise injection}, along with a general approach for \emph{indirect ADM} when direct control is not available.

\subsubsection{Fingerprinting} \label{sec:adm_fingerprinting}
\begin{figure}
    \centering
    \includegraphics[width=0.9\columnwidth]{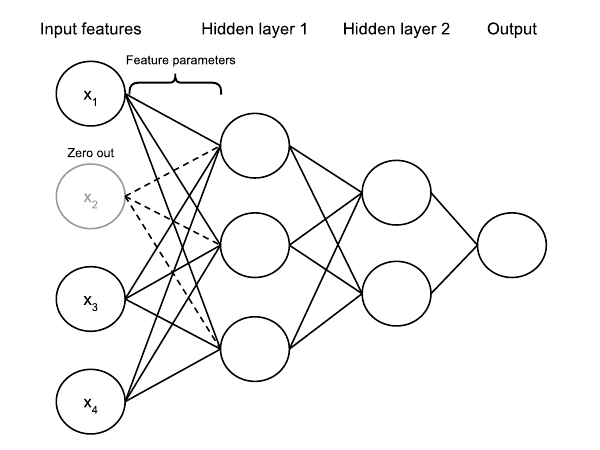}
    \caption{Example of the fingerprinting method for a 2-layer neural net. The feature parameters consist of all connections between the inputs and the first hidden layer. Feature $x_2$ is zero-ed out, causing all feature weights corresponding to $x_2$ (dashed lines) to have 0 gradient during backpropagation.}
    \label{fig:adm}
\end{figure}

In our first ADM method, the server modifies the item features to influence the training process such that the user's local update for any feature parameter can be deterministically zero or non-zero, thus acting as a form of signal or identifier for the server.
Consider any single model parameter $\theta_j$ that is in direct contact with at least one feature of the training items.
For example, if the model is linear or logistic regression, then any non-bias weight is one such parameter.
Similarly, if the model is a neural network, then any non-bias weight in the first layer of the neural net will satisfy the criterion, while weights in other layers are only indirectly involved and thus do not fit the criterion.
We refer to parameters that are in direct contact with the training features as \emph{feature parameters} (see Figure \ref{fig:adm}).

Let $F_{\theta_j}$ denote the set of features that $\theta_j$ is in direct contact with, and let $\nabla_{\theta_j}$ be the user's local update to $\theta_j$ via learning algorithm $g$.
In order for $\nabla_{\theta_j}$ to be non-trivial (i.e., non-zero), the values of the features in $F_{\theta_j}$ must also be non-trivial.
If each feature in $F_{\theta_j}$ has the same values across all items, for example, then the model will not learn anything meaningful about the features since they do not contribute meaningfully to the model predictions.
Thus, the server can deterministically control whether $\theta_j$ is updated or not by setting the values of the features in $F_{\theta_j}$ appropriately.
More specifically, to make $\nabla_{\theta_j} \approx 0$, the server can simply set each feature in $F_{\theta_j}$ to exactly or near 0, while to make $\nabla_{\theta_j} \neq 0$, the server can simply keep the original feature values.
By exclusively assigning certain features to specific items, the result of interacting with those items can be more easily determined from the model parameters updates.
Hence, we call this technique \emph{fingerprinting}.

\subsubsection{Noise Injection} \label{sec:adm_noise_injection}

Although the fingerprinting idea with zeros described above is easy to demonstrate, it is not necessarily the best ADM method.
In terms of execution, we would need to know how to assign features to each item and which features to set to zero such that the effect of interacting with one item can be isolated from all other items.
In terms of detectability, the presence of zeros can be easily checked for, thus making the manipulation too obvious.
In terms of performance, having too many zero-ed features can potentially reduce the magnitude of the gradient updates and consequently increase the difficulty of matching the gradients in the reconstruction.
Through our experiments with gradient-based federated OLTR algorithms, we find that the reconstruction is consistently successful when the feature values are replaced completely with random noise (e.g. Gaussian noise with a diagonal covariance matrix or uniform noise), especially in the case of neural net models.
This phenomenon can be attributed to the following reasons: 1) each item's impact on the final gradients will be more ``visible'' due to the uniqueness of the item's features, thus facilitating the reconstruction; and 2) neural nets have a great capacity for learning from random noise and can even memorize them~\cite{arpit2017memorization}.
Using random noise also has two major advantages compared to the fingerprinting method: firstly, we do not need any complicated procedure to generate noise, and secondly, it is less detectable than having completely zero-ed out features. We call this method \emph{noise injection}.

\subsubsection{Indirect Manipulation}

In certain scenarios, the FL server cannot arbitrarily alter the training feature values however it wants.
While lacking direct control, the server can still indirectly influence the extracted features by modifying the representation of the items.
We focus on the case where the features of the items are extracted locally by the users using a fixed procedure agreed upon by all participants.
For example, in an image recommendation system, the server can only present images to users, and the image features can only be extracted using some pre-trained computer vision model for object classification.
However, the server can still execute the noise injection method by adding small perturbations to the images such that the extracted features closely resemble the target noise~\cite{sabour2016advrep}.
The success of this method highly depends on how pliable the feature extractor is.
Our experiments with image-based ranking show that existing computer vision models are highly susceptible to our adversarial manipulation, which leads to improved reconstruction performance (Section \ref{sec:eval_results_image}).
We further enhance the reconstruction success by integrating the fingerprinting technique (Section \ref{sec:adm_img_method}).

To summarize, we present the high-level pseudocode for \name with ADM in Algorithm \ref{alg:raifle}.
Details about specific ADM implementations can be found in Section \ref{sec:adm_img_method} and \ref{sec:eval_setup}.

\begin{algorithm}
 \caption{\name with ADM via noise injection}
 \label{alg:raifle}
 \begin{algorithmic}
 \renewcommand{\algorithmicrequire}{\textbf{Input:}}
 \renewcommand{\algorithmicensure}{\textbf{Output:}}
 \REQUIRE Learning algorithm $g$, global FL model parameters $\theta$, representations of items $\mathcal{X}$
 \ENSURE Reconstructed interactions $\mathcal{I}'$
 \STATE \textit{// ADM stage}
 \STATE Determine the optimal ADM noise distribution via search
 \STATE $\mathcal{X}' \gets$ Modify $\mathcal{X}$ to induce the desired noise in training features
 \STATE \textit{// FL Stage}
 \STATE Send $\theta$ and $\mathcal{X}'$ to a user for FL training
 \STATE $\hat{\theta} \gets$ Updated model parameters from user
 \STATE \textit{// Reconstruction stage}
 \STATE $\mathcal{I}' \gets$ Random initial interactions 
 \WHILE{termination conditions are not met}
    \STATE $\theta' \gets g(\mathcal{X}', \theta, \mathcal{I}')$
    \STATE $\nabla_{\mathcal{I'}} \gets $ Autodiff $\mathcal{L}_{atk}(\hat{\theta}, \theta')$ w.r.t.\ $\mathcal{I'}$
    \STATE Update $\mathcal{I}'$ using $\nabla_{\mathcal{I'}}$
 \ENDWHILE
 \RETURN $\mathcal{I}'$
 \end{algorithmic}
 \end{algorithm}

%% file: impl_details.tex
\section{Implementation Details} \label{sec:impl_details}
We describe important implementation details for \name, including how to enable automatic differentiation and choices of optimization algorithms and loss functions.

\subsection{Enabling Differentiation w.r.t.\ Interactions} \label{raifle:diff}
Typically, user interactions in RS/OLTR are treated as discrete values (e.g., clicked or not clicked).
Furthermore, certain RS/OLTR algorithms might not be immediately differentiable w.r.t.\ the interactions $\mathcal{I}'$.
To enable \name in these cases, the server needs to modify the learning algorithm $g$.
First, the optimization algorithms employed by $g$ must be differentiable.
Furthermore, $g$ needs to treat discrete interaction values as continuous ``degrees'' of interactions, with any non-differentiable usage of the interactions to be replaced with differentiable operations.
We outline some general methods to modify representative RS/OLTR learning paradigms~\cite{liu2009ltr}:

\textbf{Pointwise:} If $\mathcal{L}_{FL}$ is not differentiable w.r.t.\ to $\mathcal{I'}$ (e.g. 0-1 classification loss), change it to a differentiable one such as $L_2$ loss.
In fact, if $\mathcal{L}_{FL}$ is $L_2$ loss, then we can show that $\nabla_{\mathcal{I'}}^2 g = \mathbf{0}$ (see Appendix \ref{apd:soln}), which implies that the attack is convex (Theorem \ref{thm:convexity}).

\textbf{Pairwise:} Instead of only including item pairs in which one item is ``preferred'' over the other (e.g. pairs of click and no click), $\mathcal{L}_{FL}$ can assign a weight to each pair based on how large the difference in preference is, then calculate the weighted sum of the losses of all pairs (see example in Section \ref{sec:eval_oltr_models}).
While this can enable once-differentiation, the Hessian of $g$ will likely not be equal to $\mathbf{0}$ since we now consider how user interactions ``interact'' with one another.

\textbf{Listwise:} There is a wide variety of listwise losses developed in the learning-to-rank literature~\cite{liu2009ltr}, some of which are naturally differentiable w.r.t.\ user interactions, while for many others the relationship is not immediately clear without redefining the loss formulas.
As in the case of pairwise losses, even though once-differentiation might be possible, $\nabla_{\mathcal{I'}}^2 g$ might not be equal to $\mathbf{0}$.
Due to the vast assortments of listwise methods and the lack of federated listwise algorithms, we leave the investigation of listwise attacks for future research.

\subsection{Optimization Algorithms and Losses}
There are various options for the loss function $\mathcal{L}$ and the optimization algorithm of \name.
Previous research in gradient inversion attacks has often used the $L_2$ loss combined with the L-BFGS optimizer~\cite{zhu2019dlg} or the cosine distance loss with the Adam optimizer~\cite{geiping2020}.
We used the $L_2$ loss with the default L-BFGS and Adam optimizers from PyTorch~\cite{pytorch2019} in our experiments, and found that while both optimizers can yield good results, PyTorch's L-BFGS is typically faster but needs to have its termination criteria appropriately set based on the magnitude of the loss.
As such, we used L-BFGS for the non-image scenarios since it works without much tuning and Adam for the image scenarios since it yields better results than un-tuned L-BFGS. With regards to the loss, $L_2$ works well in all of our scenarios, but the loss values may need to be scaled up depending on the learning rate used in the local FL training for the L-BFGS optimizer to work properly.
We simply divided the input gradients by the local FL learning rate in this case.

\subsection{ADM with Noise Injection} \label{sec:adm_img_method}

The distributional characteristics of the noise used in the ADM noise injection method (Section \ref{sec:adm_noise_injection}) need to be set appropriately to maximize reconstruction success.
The FL server can determine this empirically by simulating RAIFLE with its own data and fake interactions and performing a search of the noise's distributional parameters.
In our experiments, we performed a simple grid search and found that Gaussian noise with mean 0 and standard deviation $0.1$ and $4.0$ consistently yields good results for the tabular and image LTR experiments, respectively.
Other zeroth-order search algorithms such as evolution strategies~\cite{salimans2017es} can also be employed to find the most suitable noise distribution.

\begin{figure}
    \centering
    \includegraphics[width=0.9\columnwidth]{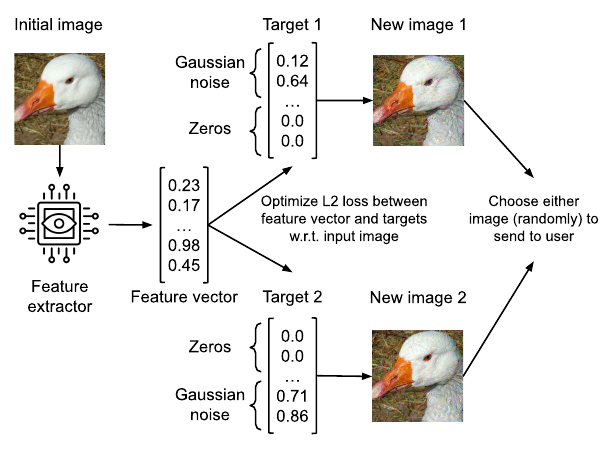}
    \caption{Diagram of our partitioned noise injection ADM method for images. The FL server prepares two manipulated versions of an image by matching the image's extracted features to two different target noise vectors.}
    \label{fig:adm_img}
\end{figure}

In the case of image-based OLTR with a pre-trained feature extractor, we used the default Adam optimizer with a learning rate of $0.01$ and 500 epochs to optimize the image inputs such that the L2 loss between the features extracted and a target Gaussian noise vector is minimized.
As the last layer in the tested vision models is the ReLU activation which only outputs non-negative values, we clip the target noise to have a minimum of 0.
Inspired by the fingerprinting technique, we also partitioned the target noise in half to create two different target vectors, with the first vector having the first half zero-ed out and the second vector having the second half zero-ed out (Figure \ref{fig:adm_img}).
Using the two resulting manipulated images, the FL server can now randomly choose which version to send to each user.
This \emph{partitioned noise injection} method allows us to minimize the loss much better than with a single noise target, as higher numbers of extracted features generally make the optimization problem more difficult.

\subsection{Software}

We use PyTorch~\cite{pytorch2019} to implement \name and the FL models for our experiments.
We use IBM's Diffprivlib Python library~\cite{diffprivlib} for the DP noise as it supports the Analytical Gaussian mechanism~\cite{balle2018} for privacy budget $\varepsilon > 1$.
For adversarial perturbation attacks, we use the Foolbox library~\cite{rauber2017foolbox, rauber2017foolboxnative}.
For differentiable optimizers, we use the Torchopt library~\cite{torchopt}.

%% file: evaluation_setup.tex
\section{Experiment Setup} \label{sec:eval_setup}
Here, we describe our experiment setup for three evaluation scenarios: federated RS, federated OLTR with non-image data, and federated OLTR with image-based data.

\subsection{Federated RS}

\begin{algorithm}[t]
 \caption{\name for Federated RS (No ADM)}
 \label{alg:raifle2}
 \begin{algorithmic}[1]
 \renewcommand{\algorithmicrequire}{\textbf{Input:}}
 \renewcommand{\algorithmicensure}{\textbf{Output:}}
 \REQUIRE Learning algorithm $g$, global model parameters $\theta$, item embeddings $\mathcal{X}$, user's updated item embeddings $\hat{\mathcal{X}}$, user's updated model parameters $\hat{\theta}$
 \ENSURE Reconstructed interactions $\mathcal{I}'$, user embedding $e'_u$
 \STATE $\mathcal{I}' \gets$ Random initial interactions
 \STATE $e'_u \gets$ Random initial user embedding
 \WHILE{termination conditions are not met}
    \STATE $\mathcal{X}', \theta' \gets g(\mathcal{X}, \theta, e'_u, \mathcal{I}')$
    \STATE $\nabla_{\mathcal{I'}}, \nabla_{e'_u} \gets $ Autodiff $\mathcal{L}_{atk}(\hat{\theta}, \theta') + \mathcal{L}'_{atk}(\hat{\mathcal{X}}, \mathcal{X}')$ wrt\ $\mathcal{I'}$, $e'_u$
    \STATE Update $\mathcal{I}'$ and $e'_u$ using $\nabla_{\mathcal{I'}}$ and $\nabla_{e'_u}$
 \ENDWHILE
 \RETURN $\mathcal{I}'$ and $e'_u$
 \end{algorithmic}
\end{algorithm}

\subsubsection{Scenario}
We focus on the collaborative filtering recommendation paradigm, specifically the FNCF algorithm~\cite{perifanis2022fncf}.
As mentioned in Section \ref{sec:fedrs}, users in FNCF keep a private user embedding $e_u$ that is not known to the FL server.
This private embedding together with the public item embeddings $\mathcal{X}$ (known to the server) form the inputs to the global FL model to produce a personalized ranking score for each recommendation item.
In addition to the learned model parameters $\hat{\theta}$, users also share the updated item embeddings $\hat{\mathcal{X}}$ with the server.
The dimension of both the private and public embedding is 64.
The FL model is a 3-layer neural net with 128, 64, and 32 hidden units and ReLU activation.
For each user, we randomize the model parameters and user embedding, then train using the Adam optimizer for 20 epochs with a learning rate of 0.001.
These choices are similar to our baseline (Section \ref{sec:raifle_rs_baseline}).

\subsubsection{Data}
We use MovieLens-100K~\cite{movielens2015}, a widely used dataset in the RS literature, and the Steam-200K dataset~\cite{steam200k} in the IMIA paper (Table \ref{tab:frs_data}).
To create the interaction labels for our experiment, we binarize the interactions between each user and each recommendation item (1 if interacted, 0 if not).
For the local FL training of each user, we randomly sampled non-interacted items with a ratio of 4:1 to interacted ones.
\begin{table}[h]
    \footnotesize
    \centering
    \caption{Some statistics on federated RS datasets}
    \begin{tabular}{cccc}
        \toprule
        Dataset & \# of users & \# of items  & \makecell{Avg. interactions\\ per user} \\
        \midrule
        MovieLens-100K & 943 & 1,682 & 106.0 \\
        Steam-200K & 12,393 & 5,155 & 10.4 \\
        \bottomrule
    \end{tabular}
    \label{tab:frs_data}
\end{table}

\subsubsection{Attack} \label{sec:raifle_rs}
To account for the use of private user embedding and public item embeddings, we modify Algorithm \ref{alg:raifle} to perform joint optimization of user interactions and private user embedding (Algorithm \ref{alg:raifle2}), with the objective function being the sum of the model parameters loss and the item embedding loss (i.e., average $L_2$ distance between each pair of original and updated item embedding).
Note that we do not apply ADM in this scenario since the server receives an embedding for each item interacted with, which allows for (almost) perfect reconstruction of user interactions (Section \ref{sec:evaluation}).

\subsubsection{Baseline} \label{sec:raifle_rs_baseline}
We use the IMIA method~\cite{yuan2023imia} as the comparison baseline for this scenario as it is the only work (to our knowledge) that attempts to reconstruct user interaction in the federated RS scenario.
Note that IMIA is a stochastic heuristics search that does not make use of the gradient information and is limited to federated RS only.
We only compare the FNCF algorithm as the federated graph recommendation approach tested in the IMIA paper does not fully specify how to protect the privacy of the graph.
We also evaluate against IMIA's proposed regularization-based defense, which penalizes the distance between the updated item embeddings and the global embeddings using the $L_1$ loss.

\subsection{Federated OLTR (non-image)} \label{sec:experiment_foltr}

\subsubsection{Scenario}
We target the FPDGD algorithm~\cite{wang2021fpdgd} which is a pairwise method to OLTR.
The FL server directly shares the training features with the users.
Users perform multi-batch stochastic gradient descent (SGD) locally, where each SGD batch corresponds to one query and the associated items (no more than 10, sampled according to the scores of the ranking model similar to the FPDGD paper) and the local learning rate is $0.1$.
For the global FL model, we use a linear regression model and various neural nets with an increasing number of hidden units (followed by ReLU activation) to check the effect of having more parameters on reconstruction.

\subsubsection{Data}
We use the LETOR 4.0 dataset and the MSLR dataset used in the FPDGD paper, specifically the training set of the first fold of MQ2007 and MQ2008 in LETOR and of MSLR-WEB10K in MSLR~\cite{letor2013} (Table \ref{tab:data}).
We preprocessed the features in MSLR by normalizing them to have mean 0.0 and standard deviation 1.0.
We simulated the user clicks using the ``navigational'' and ``information'' click chain model following prior work on federated OLTR~\cite{kharitonov2019, wang2021fpdgd}.
\begin{table}[h]
    \footnotesize
    \centering
    \caption{Some statistics on our non-image FOLTR datasets}
    \begin{tabular}{cccc}
        \toprule
        Name & Features & Queries & \makecell{Avg. \# of items\\ per query} \\
        \midrule
        MQ2008 & 46 & 471 & 20.4 \\
        MQ2007 & 46 & 1017 & 41.5 \\
        MSLR-WEB10K & 136 & 6000 & 120.5 \\
        \bottomrule
    \end{tabular}
    \label{tab:data}
\end{table}

\subsubsection{Attack} \label{sec:eval_oltr_models}
We employ \name with the noise injection ADM method.
The FL server in this case will replace all of the training features with random noise drawn from the Gaussian distribution with mean 0.0 and standard deviation 0.1.
To make FPDGD be differentiable w.r.t.\ the interactions, we utilize the technique described in Section \ref{raifle:diff} for pairwise loss, specifically using $\mathcal{I} (1 - \mathcal{I}^T)$ as the weight matrix.

\subsubsection{Baseline}
To our knowledge, no other work has attacked the federated OLTR scenario.
As such, we compare our attack with the regular gradient inversion method (Section \ref{sec:grad_inv}), which can also be viewed as \name without ADM.

\subsection{Federated OLTR (image-based)} \label{sec:eval_img_setup}

\subsubsection{Scenario}
We simulate an image ranking scenario, where the FL server sends images to users who then ``upvote'' and ``downvote'' the images.
The local FL training is done by first extracting features from the images via an object classification vision model and then performing 5 epochs of gradient descent with a learning rate of $0.01$ on a linear or neural ranking model to minimize the pointwise ranking loss (Section \ref{raifle:diff}).
Compared to the previous scenarios, this setting is more difficult as the server no longer has direct control of the training features.
Furthermore, any manipulation should be visually subtle to avoid being detected by human eyes.
We experiment with several distinct computer vision models pre-trained on ImageNet to test the generalizability of our indirect noise injection technique across different architectures and feature dimensions (see Table \ref{tab:cv_models}).
All of these models have a small number of parameters yet high accuracy on ImageNet, which makes them ideal for the FL setting where computation resources for deep learning can be constrained by the available hardware.
We extract the representations from the last layer before the final classification layer as the ranking features.
\begin{table}
    \centering
    \footnotesize
    \caption{Some statistics on tested computer vision models}
    \begin{tabular}{ccc}
        \toprule
        \thead{Model} & \thead{Number of\\ parameters} & \thead{Dimension of\\ extracted features} \\
        \midrule
        ResNet18~\cite{he2016resnet} & 11.7 mil & 512 \\
        RegNet Y 800MF~\cite{radosavovic2020regnet} & 6.4 mil & 784\\
        DenseNet121~\cite{huang2017densenet} & 8.0 mil & 1024 \\
        MNasNet 1.3~\cite{tan2019mnas} & 6.3 mil & 1280 \\
        \bottomrule
    \end{tabular}
    \label{tab:cv_models}
\end{table}

\begin{figure}[t]
    \footnotesize
    \centering
    \renewcommand{\arraystretch}{0}
    \begin{tabular}{c@{\hspace{1.5em}\vspace{0.2em}}c@{\hspace{0.2em}}c@{\hspace{0.2em}}c@{\hspace{0.2em}}c}
        Original & \adjustimage{width=1.25cm,valign=m}{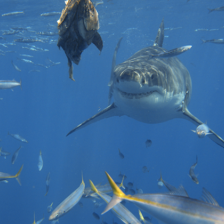} & \adjustimage{width=1.25cm,valign=m}{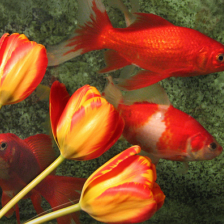} & \adjustimage{width=1.25cm,valign=m}{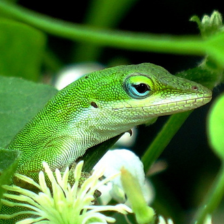} & \adjustimage{width=1.25cm,valign=m}{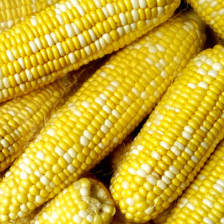}\\
        ResNet18 & \adjustimage{width=1.25cm,valign=m}{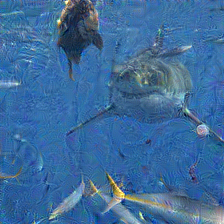} & \adjustimage{width=1.25cm,valign=m}{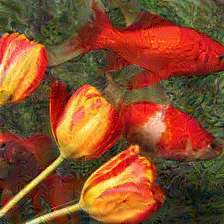} & \adjustimage{width=1.25cm,valign=m}{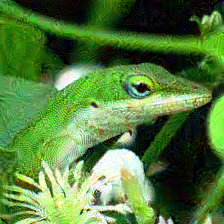} & \adjustimage{width=1.25cm,valign=m}{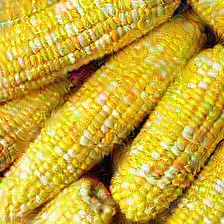}\\
        RegNet Y 800MF & \adjustimage{width=1.25cm,valign=m}{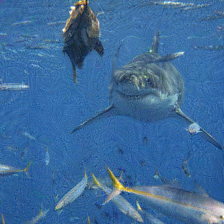} & \adjustimage{width=1.25cm,valign=m}{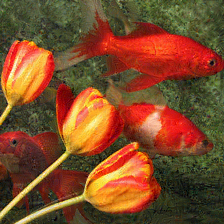} & \adjustimage{width=1.25cm,valign=m}{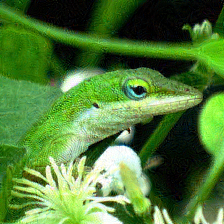} & \adjustimage{width=1.25cm,valign=m}{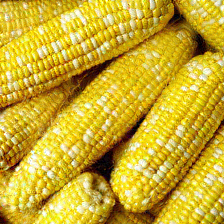}\\
        DenseNet121 & \adjustimage{width=1.25cm,valign=m}{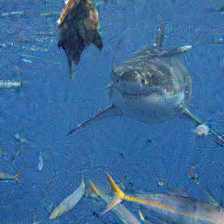} & \adjustimage{width=1.25cm,valign=m}{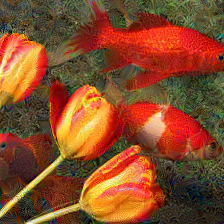} & \adjustimage{width=1.25cm,valign=m}{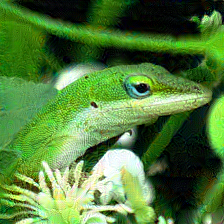} & \adjustimage{width=1.25cm,valign=m}{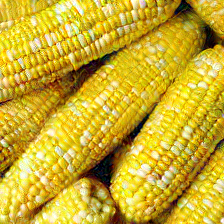}\\
        MNasNet 1.3 & \adjustimage{width=1.25cm,valign=m}{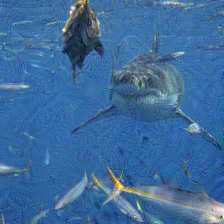} & \adjustimage{width=1.25cm,valign=m}{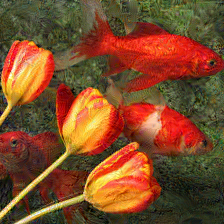} & \adjustimage{width=1.25cm,valign=m}{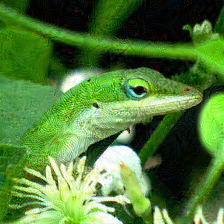} & \adjustimage{width=1.25cm,valign=m}{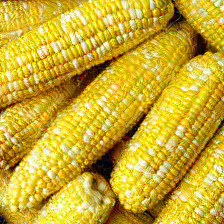}\\
    \end{tabular}
    \caption{Examples of original and manipulated images from ImageNet for different vision models. Some artifacts are visible but subtle.}
    \label{fig:adm_images}
\end{figure}

\subsubsection{Data}
We use the ImageNet-1K (2012) dataset~\cite{russakovsky2015imagenet}, particularly the validation split which consists of 50,000 images equally spread across 1,000 different classes.
Each image is preprocessed using the default transformation in PyTorch's pre-trained vision models (i.e., resized and centrally cropped to size $224 \times 224$ then normalized via ImageNet standardization).

\subsubsection{Attack}
We use the partitioned noise injection method (Section \ref{sec:adm_img_method}) to manipulate each image for each target vision model.
After performing ADM on an image, we undo the ImageNet normalization, convert the pixel values to 8-bit unsigned integers, and save the image in PNG format (about 147kB each).
This results in a negligible difference of about $10^{-5}$ to $10^{-4}$ mean $L_2$ error in the preprocessed pixel values before and after saving (Figure \ref{fig:adm_images}).
To create user interactions, we randomly select each image as interacted with probability $0.5$.
To perform the reconstruction attack, we use the Adam optimizer with a learning rate of $0.1$ running for 300 epochs.

\subsubsection{Baseline}
Similar to the non-image federated OLTR scenario, we use the vanilla gradient inversion without ADM as the baseline.
Additionally, we test the Fast Gradient Sign Method (FGSM)~\cite{goodfellow2014fgsm} which perturbs the images to cause the models to misclassify by adding $\epsilon$ times the signs of the gradients of the classification loss w.r.t. the pixels ($\epsilon = 0.1$).

%% file: evaluation_results.tex
\section{Evaluation Results} \label{sec:evaluation}

In this section, we present our evaluation results for the scenarios described in Section \ref{sec:eval_setup}.
We primarily report the \emph{area under the curve} of the receiver operating characteristic (AUC)~\cite{rocauc} since we do not want to rely on a fixed threshold for the classification and we equally care about positive and negative interactions.
Note that randomly guessing will result in an AUC of $\approx 0.5$.
We summarize some highlights of our findings as follows:
\begin{itemize}
    \item \name consistently outperforms baselines such as IMIA for federated RS and gradient inversion for both federated OLTR scenarios, achieving 0.8-1.0 AUC in most cases.
    \item \name achieves superior performance even when the server does not have direct control of the training features, particularly in the case of image-based federated OLTR.
    \item \name scales better than other techniques when the ratio of items to features increases.
    \item \name can better utilize an increasing number of model parameters than other techniques.
\end{itemize}

\subsection{Federated RS} \label{sec:eval_results_rec}
Table \ref{tab:frs_results} presents \name's reconstruction performance on the FNCF algorithm for the Movie Lens-100K and Steam-200K datasets.
We executed \name on each user exactly once.
\name is able to achieve near-perfect reconstruction in this scenario even when ADM is not used.
This can be explained by the fact that users share with the FL server an item embedding for each item, thus allowing the effect of each interaction to be easily determined.
In fact, it is not necessary to use the model parameters for the reconstruction, as the item embeddings by themselves are sufficient.

Compared to the IMIA method, \name has significantly better performance, achieving $>0.9$ F1 scores on both MovieLens-100K and Steam-200K (Table \ref{tab:raifle_vs_imia}).
With IMIA's proposed defense applied (L1 regularization term set to $1.0$), \name still manages to achieve higher F1 scores than IMIA.
Our results thus demonstrate the superiority of gradient-based optimization over a heuristic search like IMIA.
Note that we can only compare the two methods using the F1 scores since IMIA only outputs a hard label.
\begin{table}[h!]
    \centering
    \caption{\name's AUC on MovieLens-100K and Steam-200K}
    \footnotesize
    \begin{tabular}{cccc}
        \toprule
        Dataset & Mean & Median & \thead{Standard\\ deviation} \\
        \midrule
        MovieLens-100K & 0.998 & 1.000 & 0.003 \\
        Steam-200K & 0.960 & 1.000 & 0.190 \\
        \bottomrule
    \end{tabular}
    \label{tab:frs_results}
\end{table}
\begin{table}[h!]
    \centering
    \footnotesize
    \caption{Avg. F1 scores (threshold 0.5) for \name and IMIA (from \cite{yuan2023imia}). \name outperforms IMIA in both datasets by $>0.1$ with IMIA defense and $>0.3$ without.}
    \begin{tabular}{cccc}
        \toprule
        Method & IMIA Defense & MovieLens-100K & Steam-200K \\
        \midrule
        IMIA & No & 0.593 & 0.671 \\
        \name & No & \textbf{0.983}  & \textbf{0.923} \\
        \midrule        
        IMIA & Yes & 0.215 & 0.206 \\
        \name & Yes & \textbf{0.382} & \textbf{0.316} \\
        \bottomrule
    \end{tabular}
    \label{tab:raifle_vs_imia}
\end{table}

\subsection{Federated OLTR (non-image)} \label{sec:eval_results_foltr}

\begin{table}[!ht]
    \scriptsize
    \centering
    \caption{Mean reconstruction AUC for FPDGD on MQ2007 and MSLR-WEB10K. \name outperforms vanilla gradient inversion in all scenarios by a significant margin.}
    \begin{tabular}{cccccccc}
        \multicolumn{8}{c}{MQ2007} \\
        \toprule
        \multirow{2}{*}{\makecell{Model}} & \multirow{2}{*}{ADM} & \multicolumn{3}{c}{Informational} & \multicolumn{3}{c}{Navigational} \\
        \cmidrule(lr){3-5}\cmidrule(lr){6-8}
         &  & 4 & 8 & 16 & 4 & 8 & 16 \\
        \midrule
        \multirow{2}{*}{\makecell{Linear}} & None & 0.87 & 0.76 & 0.66 & 0.95 & 0.84 & 0.71 \\
         & \name & \textbf{1.00} & \textbf{0.98} & \textbf{0.82} & \textbf{1.00} & \textbf{0.98} & \textbf{0.88} \\
        \midrule
        \multirow{2}{*}{\makecell{Neural\\(4 hid. units)}} & None & 0.78 & 0.66 & 0.57 & 0.86 & 0.70 & 0.60 \\
         & \name & \textbf{1.00} & \textbf{0.99} & \textbf{0.95} & \textbf{1.00} & \textbf{1.00} & \textbf{0.98} \\
        \midrule
        \multirow{2}{*}{\makecell{Neural\\(8 hid. units)}} & None & 0.77 & 0.64 & 0.56 & 0.86 & 0.68 & 0.59 \\
         & \name & \textbf{1.00} & \textbf{1.00} & \textbf{0.99} & \textbf{1.00} & \textbf{1.00} & \textbf{1.00} \\
        \midrule
        \multirow{2}{*}{\makecell{Neural\\(16 hid. units)}} & None & 0.75 & 0.60 & 0.53 & 0.82 & 0.61 & 0.55 \\
         & \name & \textbf{1.00} & \textbf{1.00} & \textbf{1.00} & \textbf{1.00} & \textbf{1.00} & \textbf{1.00} \\
        \midrule
        \midrule
        \multicolumn{8}{c}{MSLR-WEB10K} \\
        \multirow{2}{*}{\makecell{Model}} & \multirow{2}{*}{ADM} & \multicolumn{3}{c}{Informational} & \multicolumn{3}{c}{Navigational} \\
        \cmidrule(lr){3-5}\cmidrule(lr){6-8}
         &  & 12 & 24 & 48 & 12 & 24 & 48 \\
        \midrule
        \multirow{2}{*}{\makecell{Linear}} & None & 0.54 & 0.51 & 0.50 & 0.55 & 0.51 & 0.51 \\
         & \name & \textbf{1.00} & \textbf{0.98} & \textbf{0.80} & \textbf{1.00} & \textbf{0.97} & \textbf{0.86} \\
        \midrule
        \multirow{2}{*}{\makecell{Neural\\(4 hid. units)}} & None & 0.52 & 0.50 & 0.50 & 0.52 & 0.51 & 0.50 \\
         & \name & \textbf{0.98} & \textbf{0.95} & \textbf{0.87} & \textbf{0.99} & \textbf{0.97} & \textbf{0.93} \\
        \midrule
        \multirow{2}{*}{\makecell{Neural\\(8 hid. units)}} & None & 0.51 & 0.50 & 0.50 & 0.51 & 0.50 & 0.50 \\
         & \name & \textbf{0.99} & \textbf{0.96} & \textbf{0.91} & \textbf{1.00} & \textbf{0.99} & \textbf{0.96} \\
        \midrule
        \multirow{2}{*}{\makecell{Neural\\(16 hid. units)}} & None & 0.51 & 0.50 & 0.50 & 0.51 & 0.50 & 0.51 \\
         & \name & \textbf{0.97} & \textbf{0.94} & \textbf{0.91} & \textbf{0.99} & \textbf{0.98} & \textbf{0.96} \\
        \bottomrule
    \end{tabular}
    \label{tab:foltr_results}
\end{table}

Table \ref{tab:foltr_results} showcases \name's performance for the FPDGD algorithm on the MQ2007 and MSLR-WEB10K dataset (we omit MQ2008 since its result is very similar to MQ2007).
We tested each combination of ranking model and click model with 4, 8, and 16 total queries for the multibatch local training on MQ2007, and 12, 24, and 48 queries on MSLR-WEB10K.
These numbers of queries are intentionally chosen to scale with the number of available features in each dataset (e.g. 4 queries and 12 queries result in nearly the same number of items as the number of features in MQ2007 and MSLR-WEB10K, respectively, while 8, 16 and 24, 48 are double and quadruple).
We can see that \name with our noise injection ADM method has much better AUC than \name without ADM (i.e., vanilla gradient inversion), achieving $>0.9$ AUC in all scenarios except for the linear ranker with 16 queries setting, which is also the most difficult since it has the least number of model parameters and the most number of items.
Particularly, on the MSLR-WEB10K dataset, the reconstruction without ADM is barely better than random guessing.
Furthermore, \name with ADM is more capable of utilizing the increasing number of model parameters to attain better reconstruction results than without ADM.

\begin{table*}[!ht]
    \footnotesize
    \centering
    \caption{Mean reconstruction AUC (rounded to 3rd decimal place) for federated OLTR on image-based data. The 1x, 2x, and 4x underneath the ranking model type refer to the ratio of images to feature dimensions. Each configuration was run 200 times with randomly sampled model parameters and interactions. \name outperforms baselines in most scenarios.}
    \begin{tabular}{cccccccccccccc}
        \toprule
        \multirow{2}{*}{\makecell{Vision Model}} & \multirow{2}{*}{ADM} & \multicolumn{3}{c}{Linear} & \multicolumn{3}{c}{Neural (2 hidden units)} & \multicolumn{3}{c}{Neural (4 hidden units)} & \multicolumn{3}{c}{Neural (8 hidden units)} \\
        \cmidrule(lr){3-5}\cmidrule(lr){6-8}\cmidrule(lr){9-11}\cmidrule(lr){12-14}
         &  & 1x & 2x & 4x & 1x & 2x & 4x & 1x & 2x & 4x & 1x & 2x & 4x \\
        \midrule
        \multirow{3}{*}{\makecell{ResNet18\\(512 features)}} & None & 1.000 & 0.916 & 0.767 & 0.921 & 0.868 & 0.771 & 0.985 & 0.945 & 0.857 & 0.999 & 0.985 & 0.924 \\
         & FGSM & 1.000 & 0.915 & 0.766 & 0.914 & 0.857 & 0.759 & 0.981 & 0.934 & 0.841 & 0.998 & 0.977 & 0.906 \\
         & \name & \textbf{1.000} & \textbf{0.943} & \textbf{0.772} & \textbf{0.946} & \textbf{0.920} & \textbf{0.823} & \textbf{0.993} & \textbf{0.981} & \textbf{0.922} & \textbf{1.000} & \textbf{0.998} & \textbf{0.978} \\
        \midrule
        \multirow{3}{*}{\makecell{RegNet Y 800MF\\(784 features)}} & None & 0.999 & 0.918 & \textbf{0.772} & 0.948 & 0.913 & 0.801 & 0.991 & 0.977 & 0.891 & 1.000 & 0.994 & 0.925 \\
         & FGSM & 0.999 & 0.913 & 0.771 & 0.948 & 0.908 & 0.796 & 0.990 & 0.974 & 0.884 & 1.000 & 0.993 & 0.922 \\
         & \name & \textbf{1.000} & \textbf{0.952} & 0.767 & \textbf{0.956} & \textbf{0.932} & \textbf{0.833} & \textbf{0.993} & \textbf{0.989} & \textbf{0.938} & \textbf{1.000} & \textbf{0.999} & \textbf{0.984} \\
        \midrule
        \multirow{3}{*}{\makecell{DenseNet121\\(1024 features)}} & None & \textbf{0.933} &  \textbf{0.772} & \textbf{0.667} & 0.902 & 0.805 & 0.700 & 0.970 & 0.896 & 0.770 & 0.995 & 0.946 & 0.827 \\
         & FGSM & 0.933 & 0.771 & 0.666 & 0.891 & 0.794 & 0.691 & 0.961 & 0.884 & 0.759 & 0.992 & 0.932 & 0.810 \\
         & \name & 0.919 & 0.765 & 0.664 & \textbf{0.923} & \textbf{0.825} & \textbf{0.717} & \textbf{0.983} & \textbf{0.935} & \textbf{0.815} & \textbf{0.999} & \textbf{0.983} &  \textbf{0.904} \\
        \midrule
        \multirow{3}{*}{\makecell{MNasNet 1.3\\(1280 features)}} & None & 0.994 & 0.895 & 0.764 & 0.936 & 0.858 & 0.754 & 0.985 & 0.926 & 0.787 & 0.996 & 0.942 & 0.775 \\
         & FGSM & 0.989 & 0.885 & 0.758 & 0.930 & 0.845 & 0.742 & 0.982 & 0.913 & 0.771 & 0.994 & 0.928 & 0.755 \\
         & \name & \textbf{1.000} & \textbf{0.939} & \textbf{0.775} & \textbf{0.940} & \textbf{0.882} & \textbf{0.791} & \textbf{0.991} & \textbf{0.965} & \textbf{0.875} & \textbf{0.999} & \textbf{0.992} & \textbf{0.924} \\
        \bottomrule
    \end{tabular}
    \label{tab:image_results}
\end{table*}
\subsection{Federated OLTR (image-based)} \label{sec:eval_results_image}
Table \ref{tab:image_results} presents the mean AUC and standard deviation of \name and baselines for each vision model, ranker model (linear and neural net with 2, 4, and 8 hidden units, ReLU activation), and ratio of images to feature dimensions (1x, 2x, and 4x).
Overall, we can observe that \name consistently outperforms vanilla gradient inversion and FGSM across most settings.
The only case where \name is not consistently better is the linear ranker scenario, although \name is still within $0.02$ AUC from the best method.
As the number of images increases, the reconstruction becomes harder overall, but \name is least affected.
The FGSM method does not improve upon vanilla gradient inversion, which indicates that perturbing the images to cause misclassification does not automatically extend to our problem of reconstructing interactions.

%% file: countermeasures.tex
\section{Countermeasures} \label{sec:countermeasures}

In this section, we look at several possible countermeasures for \name, including local differential privacy (LDP) and secure aggregation (SA).
See Appendix \ref{apd:countermeasures} for more discussion.

\subsection{Local Differential Privacy}
To test the effect of LDP on \name, we experiment with the Gaussian mechanism by applying Gaussian noise to each user's local update using privacy budget $\varepsilon \in \{1, 20, 100, 500\}$ ($\delta = 10^{-8}$).
Table \ref{tab:ldp_results} shows the results of applying LDP on two example configurations from each evaluation scenario in Section \ref{sec:eval_setup} (FNCF with ML-100K and Steam-200K for federated RS; FPDGD with linear ranker, informational click model, and 16 queries on MQ 2007 and MSLR-WEB10K; neural ranker with 8 hidden units and number of items equal to number of features using ResNet18 and DenseNet121).
$\varepsilon$ is chosen to represent a wide range of privacy protections, with $\varepsilon = 1$ providing particularly strong guarantees under LDP (at the expense of utility) and $\varepsilon = 500$ representing a typical LDP budget adopted in LDP research and applications~\cite{bhowmick2019, google2023}. 
We set the sensitivity $\Delta_2 = 0.1, 0.5, 0.05$ for federated RS, FPDGD, and FOLTR with images, respectively.
We use the Analytical Gaussian formula~\cite{balle2018} for $\varepsilon > 1$ and the standard one for $\varepsilon = 1$~\cite{dwork2014}.
\begin{table}
    \centering
    \caption{Mean reconstruction AUC of some representative evaluation scenarios with LDP applied.}
    \setlength{\tabcolsep}{3pt}
    \begin{tabular}{ccccccc}
        \toprule
        Scenario & ADM & $\varepsilon\!=\!1$ & $\varepsilon\!=\!20$ & $\varepsilon\!=\!100$ & $\varepsilon\!=\!500$ & No DP \\
        \midrule
        \makecell{FNCF w/\\ ML-100K} & N/A & 0.50 & 0.52 & 0.56 & 0.74 & 1.00 \\
        \midrule
        \makecell{FNCF w/\\ Steam-200K} & N/A & 0.50 & 0.56 & 0.72 & 0.90 & 0.96 \\
        \midrule
        \multirow{2}{*}{\makecell{FPDGD w/\\MQ 2007}} & None & 0.50 & 0.54 & 0.57 & 0.58 & 0.66 \\
        & \name & 0.50 & 0.56 & 0.66 & 0.75 & 0.82 \\
        \midrule
        \multirow{2}{*}{\makecell{FPDGD w/\\MSLR10K}} & None & 0.50 & 0.50 & 0.50 & 0.50 & 0.50 \\
        & \name & 0.50 & 0.52 & 0.58 & 0.62 & 0.80 \\
        \midrule
        \multirow{2}{*}{\makecell{FOLTR w/\\ResNet18}} & None & 0.50 & 0.52 & 0.55 & 0.62 & 1.00 \\
        & \name & 0.50 & 0.52 & 0.55 & 0.63 & 1.00 \\
        \midrule
        \multirow{2}{*}{\makecell{FOLTR w/\\DenseNet121}} & None & 0.50 & 0.51 & 0.54 & 0.59 & 1.00 \\
        & \name & 0.50 & 0.51 & 0.53 & 0.59 & 1.00 \\
        \bottomrule
    \end{tabular}
    \label{tab:ldp_results}
\end{table}

\begin{table}
    \centering
    \caption{FPDGD's avg. test NDCG@10 vs LDP $\varepsilon$ (MSLR-10K).}
    \begin{tabular}{ccccccc}
        \toprule
        Model & Click model & $\varepsilon\!=\!1$ & $\varepsilon\!=\!20$ & $\varepsilon\!=\!100$ & $\varepsilon\!=\!500$ & No DP \\
        \midrule
        Linear & Informational & 0.228 & 0.280 & 0.298 & 0.301 & 0.315 \\
        Linear & Navigational & 0.225 & 0.294 & 0.310 & 0.309 & 0.328 \\
        Neural & Informational & 0.223 & 0.226 & 0.218 & 0.228 & 0.282 \\
        Neural & Navigational & 0.227 & 0.243 & 0.256 & 0.263 & 0.291 \\
        \bottomrule
    \end{tabular}
    \label{tab:ldp_utility}
\end{table}

Overall, we see that LDP can reduce \name's effectiveness close to random guessing, but only with sufficiently small $\varepsilon$.
At $\varepsilon=500$, while the reconstruction performance is decreased across all scenarios and methods, for FNCF, FPDGD on MQ2007, and FOLTR with images, the performance remains relatively competitive.
Non-ADM and ADM's performance in the two image-based configurations is essentially the same, but in the FPDGD case, ADM is slightly better.
This higher performance from ADM is not because ADM can overcome LDP, but rather because without ADM in those cases, the reconstruction would not even work.

Despite the capability of LDP, it can significantly reduce the utility of the learned FL model~\cite{wei2020}.
To illustrate the privacy-utility tradeoff, we simulate the FL training for the FPDGD scenario on the MSLR-WEB10K dataset with a linear model and a neural net with 16 hidden units.
The FL model is trained on a single pass of the train portion in Fold 1 of the dataset, where each user is assigned one unique query.
FL aggregation occurs for every 100 users.
We measure the normalized discounted cumulative gain for the top 10 items (NDCG@10) on the corresponding test portion.
The simulation is repeated 10 times.
Our results show that even for $\varepsilon\!=\!100$, the FL model's performance suffers a noticeable decrease (Table \ref{tab:ldp_utility}).
Thus, FL practitioners often need to use large LDP $\varepsilon>10$ to achieve practical model performance~\cite{bhowmick2019, google2023}.
Determining the right balance between privacy protection and application utility is essential to adequately defend against attacks like \name.
For more results on the privacy-utility tradeoff of FNCF and FPDGD, we refer readers to Table 4 in the IMIA paper \cite{yuan2023imia} and Table 2 in the FPDGD  paper \cite{wang2021fpdgd}.

\subsection{Secure Aggregation with DP} \label{sec:countermeasures_sa}
Due to the high utility impact of LDP, a more effective defense is to combine SA and DP to reduce the amount of noise needed while achieving a similar level of privacy~\cite{kairouz2021sa}.
Although this approach can hide the link between user updates and their identities, it is not invulnerable to our ADM technique.
The FL server can single out the local update for a target user $u_k$ from the securely aggregated result by executing our fingerprinting technique, optionally in combination with noise injection (Figure \ref{fig:bypass_sa}):
\begin{enumerate}
    \item Choose a subset $D$ of the training features.
    \item For target user $u_k$: set all features $\notin D$ to 0. (Optional: Inject noise into all features $\in D$).
    \item For all other users: set all features $\in D$ to $0$.
    \item Send the modified features to users for the FL training.
\end{enumerate}

\begin{figure}[!h]
    \begin{center}
        \footnotesize
        \begin{tabular}{c|cccccc}
            \toprule
            User & $d_1$ & $d_2$ & $d_3$ & $d_4$ & $d_5$ & $d_6$ \\
            \hline
            \vdots & \vdots & \vdots & \vdots & \vdots & \vdots & \vdots \\
            $u_{k-1}$ & 0.1 & 0.2 & 0.3 & \cellcolor{lightgray}\textbf{0.0} & \cellcolor{lightgray}\textbf{0.0} & \cellcolor{lightgray}\textbf{0.0} \\
            $\boldsymbol{u_k}$ & \cellcolor{lightgray}\textbf{0.0} & \cellcolor{lightgray}\textbf{0.0} & \cellcolor{lightgray}\textbf{0.0} & 0.4 & 0.5 & 0.6 \\
            $u_{k+1}$ & 0.1 & 0.2 & 0.3 & \cellcolor{lightgray}\textbf{0.0} & \cellcolor{lightgray}\textbf{0.0} & \cellcolor{lightgray}\textbf{0.0} \\
            \vdots & \vdots & \vdots & \vdots & \vdots & \vdots & \vdots \\
            \bottomrule
        \end{tabular}
    \end{center}
    \caption{Example of bypassing vanilla SA with 6 training features via our fingerprinting ADM technique. Target user $u_k$ has features $d_1, d_2, d_3$ set to 0 while features $d_4, d_5, d_6$ are non-zero.}
    \label{fig:bypass_sa}
\end{figure}

As described in Section \ref{sec:manipulation_approach}, the local update from user $u_k$ will be non-zero for the feature parameters in direct contact with $D$ and will be zero for all other feature parameters, while for all other users, the reverse applies.
Thus, when the local updates from all users are securely aggregated (e.g. via summation), the final result will leak $u_k$'s local update since we are only adding $0$'s from other users to $u_k$'s update.
Consequently, the server can now execute \name on user $u_k$ using the singled-out updates.

To illustrate the performance of \name against SA, we simulate SA with Gaussian LDP noise and a varying number of FL participants $n$ in the FPDGD scenario on the MQ2007 dataset.
Using the above technique with noise injection, we randomly choose one target user $u_k$ and let $D$ be the entire feature space, thus setting the gradients of all other users to $\mathbf{0}$.
Effectively, the final aggregated result is equal to:
$$\nabla_{u_k} + N(0, n \sigma_{\varepsilon}^2)$$
where $\sigma_{\varepsilon}$ is the noise scale corresponding to LDP $\varepsilon$.
From Table \ref{tab:sec_agg}, we see that compared to using LDP alone, SA with LDP can better protect user privacy against \name with a sufficiently large number of participants, but it does not completely prevent leakage.

\begin{table}
    \footnotesize
    \centering
    \caption{\name's average AUC on FPDGD for MQ2007 with SA and LDP (4 queries, informational click model)}
    \begin{tabular}{cccccc}
        \toprule
        \multirow{3}{*}{Model} & \multirow{3}{*}{$\varepsilon$} & \multicolumn{4}{c}{Number of participants} \\
        \cmidrule(lr){3-6}
        & & 10 & 100 & 500 & 1000 \\
        \midrule
        \multirow{5}{*}{Linear} & $\infty$ & 1.00 & 1.00 & 1.00 & 1.00 \\
        & 700 & 0.79 & 0.64 & 0.56 & 0.54 \\
        & 500 & 0.75 & 0.60 & 0.55 & 0.54 \\
        & 300 & 0.71 & 0.57 & 0.54 & 0.53 \\
        & 100 & 0.61 & 0.54 & 0.52 & 0.51 \\
        \midrule
        \multirow{5}{*}{\makecell{Neural\\(16 hidden units)}} & $\infty$ & 1.00 & 1.00 & 1.00 & 1.00 \\
        & 700 & 0.81 & 0.63 & 0.56 & 0.54 \\
        & 500 & 0.77 & 0.60 & 0.54 & 0.53 \\
        & 300 & 0.73 & 0.58 & 0.52 & 0.51 \\
        & 100 & 0.61 & 0.54 & 0.51 & 0.51 \\
        \bottomrule
    \end{tabular}
    \label{tab:sec_agg}
\end{table}

Note that this attack is not restricted to just a single target user.
By allocating a distinct portion of the feature space to each target user and making sure that only each such user can have non-zero values for their allocated features, the server can learn their individual update from the final aggregated result by looking at the corresponding feature parameters.
However, this also reduces the number of available features for reconstruction and will likely affect the attack performance.
If the server can choose the participants, then this attack can also be combined with Sybil-based attacks~\cite{boenisch2023sybil} to remove all DP noise from other users, effectively getting rid of SA.

\subsection{Detecting Data Manipulation}

\subsubsection{Cryptography}
To prevent the server from executing ADM, users can validate the integrity and authenticity of the items sent by the server using cryptography techniques.
For example, users can compare cryptographic hashes (i.e., checksum) of the items to ensure that their training features are the same as all of their peers (similar to the parameters validation defense for model manipulation in~\cite{pasquini2022}).
However, different users can have different items, which renders the process of cross-checking items more difficult, not to mention the need to do so in a privacy-preserving manner to avoid leaking information to other users.
Furthermore, detecting inconsistency among users would fail if the server manipulates the data in the same manner for all users.

\subsubsection{Heuristic Checks}
One possible ADM detection method is to analyze how it affects the resulting FL gradients.
To visualize the effects, we use the t-SNE method~\cite{maaten2008tsne} with two components and varying perplexity on the non-image and image-based FOLTR scenarios.
We see that ADM-impacted gradients exhibit some noticeable differences in the case of FPDGD on MSLR-WEB10K with a neural ranker (Figure \ref{fig:tsne_mslr10k}).
However, in the case of image-based FOLTR with ResNet18 using a neural ranker, we do not observe any obvious difference (Figure \ref{fig:tsne_resnet18}).
This indicates that looking at the gradients alone does not necessarily tell us if the training data has been tampered with or not.

Another method is to directly check the training data for any sign of ADM.
Detecting our fingerprinting method is straightforward: simply checking for the presence of zeroed-out (or constant) features would suffice.
Detecting noise injection, however, is more difficult: while we can try to quantify the degree of randomness in the data (e.g., via Shannon entropy), a malicious server can mask the noise, such as by mixing noise and real data via a convex combination.
As another example, our image-based attack (Section \ref{sec:adm_img_method}) can include a regularizer to further reduce the visual artifacts (Appendix \ref{apd:img_quality}).
Designing a comprehensive set of data quality checks is likely not practically feasible due to the vast possibilities of manipulation techniques and data modalities.
The difficulty of detecting active manipulation is also reflected in model manipulation attacks~\cite{boenisch2023reconstruct}.

\subsection{Minimizing Shared Information}
While current FL schemes often rely on the sharing of gradients or parameters, such high-dimensional information can be easily exploited to breach user privacy as we have empirically shown in our paper.
From our analysis of the implications of Theorem \ref{thm:convexity}, the number of shared parameters must be greater than or equal to the number of items for the reconstruction to be able to find a unique solution.
Thus, to make the reconstruction attack more difficult, one approach is to reduce the number of shared parameters or to share something else completely different.
The FOLtR-ES algorithm~\cite{kharitonov2019} is one example federated OLTR algorithm that allows users to share with the FL server a single number for the local ranker's utility (using some locally sampled model parameters).
In OLTR, such utility metric (e.g., discounted cumulative gain) is often not differentiable w.r.t. the interactions, thus preventing \name from working properly.
While such gradient-free methods can potentially deter reconstruction attacks, they often have lower utility than gradient-based methods.
Furthermore, DP noise is still necessary to guarantee theoretical privacy protection, although the amount needed might be lower than when the full model parameters are shared.

\subsection{Personalization}

Personalized FL (PFL) is an emerging FL approach that aims to learn a customized model for each user to address the issue of heterogeneous data as well as provide user-specific personalization~\cite{tan2022pfl}.
In the context of federated RS, collaborative filtering-based methods such as FNCF can be considered PFL since they learn a private user embedding for each user.
From a privacy point of view, PFL offers both opportunities and challenges: introducing a privately-learned component can potentially help users rely less on the global model, thus reducing the amount of information needed to share with the server, but at the same time, the private component does not guarantee complete privacy protection.
Our experiment with federated RS demonstrates that the FL server can still achieve excellent reconstruction performance despite the hidden FNCF user embedding.
Recently, there has been some attempt at personalizing the user's view of the data through private \emph{item} embeddings~\cite{li2024pfl} for collaborative filtering.
This strategy could potentially prevent \name since the number of unknown parameters to estimate would be higher than the number of items.
Overall, we believe PFL can be a promising approach towards privacy, although careful design is still recommended.

%% file: discussion.tex
\section{Discussion} \label{sec:discussion}
We discuss \name's characteristics, current limitations, and possible future extensions.

\subsection{Impact on FL Utility}
\begin{figure}[t]
    \centering
    \includegraphics[width=1\linewidth]{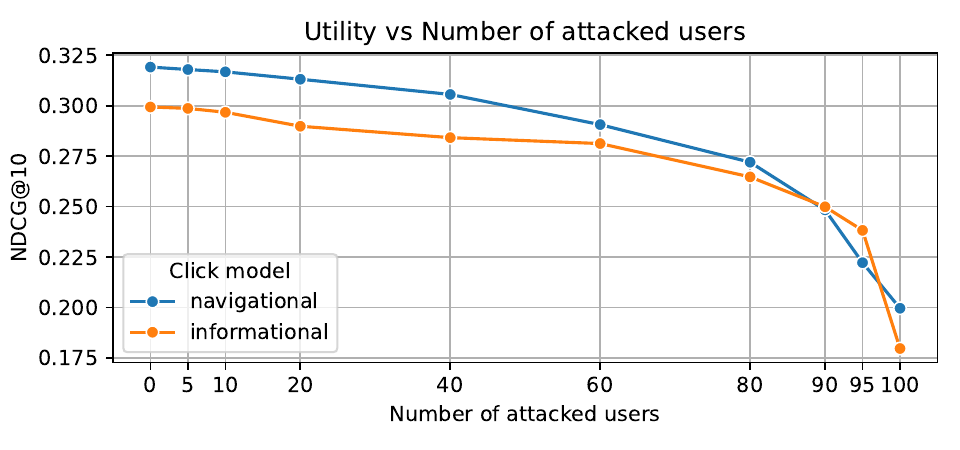}
    \vspace*{-5mm}
    \caption{Utility (test NDCG@10) vs number of attacked users (out of 100) for a linear FPDGD ranker on the MSLR-WEB10K dataset.}
    \label{fig:utility_vs_num_attacked_users}
\end{figure}

While \name requires some modifications to the local learning algorithm (Section \ref{sec:raifle_framework}), this modified version is only used by the server for performing the gradient matching optimization and is never used by real users, thus never affecting the FL model.
The use of manipulated data for ADM, however, will certainly reduce the performance of the global FL model if not handled properly.
Depending on the FL setup and the amount of control available, the server can take specific measures to limit the negative impacts of ADM.
If SA is not applied, the server can simply disregard all FL updates from targeted users since their identities are known.
Should discarding bad FL updates be infeasible, the server can choose a small number of users to apply ADM in one or a few FL rounds only instead of attacking every user in every round, thereby minimizing the low-quality contributions from targeted users.
As a demonstration, from Figure \ref{fig:utility_vs_num_attacked_users}, the test NDCG@10 for a linear FPDGD ranker trained on the MSLR-WEB10K dataset with 100 users is only reduced by $\approx$ 5-10\% even when 20-40\% of users have random item features.
If SA is applied, ADM will have to be performed for every user (Section \ref{sec:countermeasures_sa}), thus requiring the server to discard or significantly reduce the weight of the affected FL rounds' results, which is achievable if the server can control each FL round and the participants.

\subsection{Constrained Server Capabilities}

\begin{table}[t]
    \setlength{\tabcolsep}{4pt}
    \scriptsize
    \centering
    \caption{Percentage of manipulated features vs Mean reconstruction AUC for FPDGD on MQ2007 and MSLR-WEB10K}
    \begin{tabular}{cccccccccc}
    \toprule
    \multirow{2}{*}{\makecell{Dataset}} & \multirow{2}{*}{\makecell{Model}} & \multicolumn{4}{c}{Informational} & \multicolumn{4}{c}{Navigational} \\
        \cmidrule(lr){3-6}\cmidrule(lr){7-10}
    & & 0\% & 50\% & 75\% & 100\% & 0\% & 50\% & 75\% & 100\% \\
    \midrule
    \multirow{4}{*}{\makecell{MQ2007\\(16 queries)}} & Linear & 0.65 & 0.77 & 0.79 & 0.82 & 0.71 & 0.86 & 0.87 & 0.88 \\
    & Neural 4 & 0.57 & 0.72 & 0.88 & 0.95 & 0.60 & 0.79 & 0.95 & 0.98 \\
    & Neural 8 & 0.56 & 0.71 & 0.91 & 0.99 & 0.58 & 0.78 & 0.98 & 1.00 \\
    & Neural 16 & 0.53 & 0.67 & 0.90 & 1.00 & 0.55 & 0.74 & 0.98 & 1.00 \\
    \midrule
    \multirow{4}{*}{\makecell{MSLR-10K\\(48 queries)}} & Linear & 0.50 & 0.51 & 0.65 & 0.80 & 0.50 & 0.52 & 0.70 & 0.86 \\
    & Neural 4 & 0.50 & 0.50 & 0.54 & 0.87 & 0.50 & 0.50 & 0.55 & 0.93 \\
    & Neural 8 & 0.50 & 0.50 & 0.52 & 0.91 & 0.50 & 0.50 & 0.53 & 0.96 \\
    & Neural 16 & 0.50 & 0.50 & 0.51 & 0.91 & 0.51 & 0.50 & 0.51 & 0.96 \\
    \bottomrule
    \end{tabular}
    \label{tab:constrained}
\end{table}

In certain settings, the central server might be limited in its capability to manipulate the training features or even be non-existent (as in the case of decentralized federated recommendation systems; see Appendix \ref{apd:fl_variations}).
Here, we consider a constrained scenario where the server can only control a subset of the features.
We repeat the FPDGD experiment (Section \ref{sec:experiment_foltr}) but restrict to manipulating only 50\% and 75\% of all features.
Our results from the hardest settings tested for both MQ2007 and MSLR-WEB10K (without DP) indicate a positive correlation between \name's attack performance and the percentage of manipulated features (Table \ref{tab:constrained}).
However, we can observe that the impact varies depending on the dataset and the number of items, with MSLR-WEB10K incurring a much steeper reduction in mean attack AUC compared to MQ2007, essentially no better than random guessing when manipulating only 50\% of features.
We can also see that the effect of increasing the number of model parameters on attack AUC is mostly reversed when manipulating $<100$\% of features, with more features leading to weaker reconstruction performance (except for MQ2007 at 75\% manipulation).
In the case where no features can be manipulated but the presentation of the items can be modified (e.g., ordering), it might be possible for the server to exploit users' click bias w.r.t. the item positions or popularity~\cite{yuan2023imia}.
However, it is unclear how to turn this angle into an effective or efficient attack since we cannot readily leverage such discrete manipulation for optimization.

\subsection{Computational and Storage Overhead}

In our experiments, image-based manipulation is the only scenario where our attack needs significantly more computational resources.
With a single Nvidia Tesla V100 16GB GPU, it takes $\approx$ 5 minutes to manipulate each batch of 256 images for DenseNet121, the largest model tested.
Note that this step only needs to be done once since the manipulated images can be reused for different users.
Performing the gradient inversion in this scenario is much faster: only $\approx$ 20 seconds for each user with 4096 images on DenseNet121.
Considering the massive computing capabilities of adversaries like Google or Facebook, our attack is very cheap to perform, not to mention it can be executed separately from the FL protocol, thus avoiding any impact on the FL protocol's speed.
With regards to storage space, in the case of images, we would need to additionally store at least one extra manipulated image for every original image.
However, in the case of direct non-image manipulation, we can simply store a single random seed for each user so that we can recreate the attack noise vector anytime.

\subsection{Relationship to Model Manipulation and Other Attacks}
Although model manipulation~\cite{fowl2021robbing, pasquini2022, boenisch2023reconstruct} and \name's ADM method are both active attacks that lead to certain adversarial behaviors in users' local FL updates, they have several major differences in terms of techniques and applicability.
First, model manipulation depends on the presence of some specific model architectural components such as ReLU activation or CNN layers, whereas ADM does not have any explicit requirements for the FL models (other than differentiability w.r.t. interactions).
Second, achieving specific behaviors for the gradients is an explicit goal in model manipulation, but for data manipulation, the exact influence on the gradients is not always considered, as in the case of our noise injection method.
Third, model manipulation may not be possible in certain scenarios, such as when users cryptographically collaborate to choose the initial global model and cross-check the aggregated result at each FL round (unless the server inserts Sybil users~\cite{boenisch2023sybil}).
Achieving such consensus can be difficult with server-prepared data as users may have different views of the same items.

Aside from model manipulation, \name also shares some minor similarities with other ML attacks while still exhibiting noteworthy differences.
\name has a similar goal to that of gradient inversion~\cite{zhang2022surveygradinv, ovi2023gradinv} but focuses more on inverting the ``labels'' produced by the users.
It involves modifying input data like the adversarial perturbation or backdoor attacks~\cite{akhtar2021adv, bagdasaryan2020backdoor, li2024backdoor}, although its aim is not to mislead models but to extract more information from users.
\name thus can be considered a distinct blend of existing classes of ML attacks designed to attack IFL or similar scenarios.
We believe our attack set a novel research direction in ML system vulnerabilities and exploits, particularly interactive ML systems where users do not have full control over their experiences.

\subsection{Limitations and Future Work}

\name is currently restricted to FL settings wherein a central server controls the features of user interaction items, with federated recommendation and learning-to-rank as the two most prominent examples.
In this sense, \name is not as flexible as model manipulation attacks~\cite{fowl2021robbing, pasquini2022, boenisch2023reconstruct} which can be applied in more general (but still malicious) FL scenarios.
However, we argue that federated RS and OLTR are among the most important applications of FL to study given how widely-used services like search engines, e-commerce product suggestions, and entertainment recommendations all collect a vast amount of sensitive user data.
Furthermore, there have been several prominent real-world cases of large-scale recommendation data manipulation, with companies like Facebook, Amazon, and Google manipulating the new feeds or search recommendations for \emph{hundreds of thousands} of users~\cite{npr2014facebook, reuters2021amazon, ap2024google}.
These examples highlight the relevance of our strong threat model and the genuine risk of centralized entities exploiting their positions for their own benefit.

With regards to \name's design, this work mainly explores numeric feature representations only and is not yet extensible to non-differentiable item representations such as natural text.
User interactions are also currently restricted to values to which the reconstruction can be made differentiable.
As such, we intend to develop ADM techniques for more types of item modalities, including multimodal representations (e.g., image and text), and to investigate more IFL scenarios in which the interactions cannot be easily ``smoothened''.
Our proposed ADM techniques involve manipulating data only and were developed with knowledge of the underlying FL algorithm.
It is potentially possible to combine data manipulation with model manipulation~\cite{fowl2021robbing, pasquini2022, boenisch2023reconstruct} to achieve a better attack.
Automatic discovery of ADM, particularly for determining the best noise injection method, is also of special interest.
Lastly, we intend to look into more covert ADM methods by introducing a manipulation budget while retaining strong reconstruction performance.
Our preliminary results in this direction via restricting the number of manipulated features show a rather noticeable reduction in the attack's capability, which we hope to remediate.
We leave these for future research.

%% file: conclusion.tex
\section{Conclusion} \label{sec:conclusion}
This work identifies and explores the threat of reconstruction attacks in interaction-based federated learning (IFL), particularly in the context of federated recommender systems (RS) and online learning to rank (OLTR).
Our novel optimization-based \name method exploits the IFL server's knowledge and control over the interaction items to execute Adversarial Data Manipulation (ADM), a unique attack vector where the server actively manipulates the items.
We demonstrate that \name can achieve superior reconstruction performances compared to existing attacks in a variety of settings.
Our results shed light on the dire risk of inferring user interactions in IFL by malicious servers.
We discuss a variety of countermeasures against \name and ADM, including differential privacy, secure aggregation, and manipulation detection.

%% file: appendix.tex
\appendices
\section{}

\subsection{Closed-form Solution for RAIFLE} \label{apd:soln}
We present below the derivation for a closed-form solution for $\mathcal{I'}$ under certain assumptions regarding the learning algorithm $g$.

We are given items $\mathcal{X}$, a model $f$ with global model parameters $\theta$, updated model parameters $\hat{\theta}$, and learning algorithm $g$. Suppose that $\mathcal{L}_{atk}$  is the $L_2$ loss and $\mathcal{L}_{FL}$ is the pointwise $L_2$. We further assume that:
\begin{equation} \label{eq:apd1}
    \begin{aligned}
        g(\mathcal{X}, \theta, \mathcal{I}')
        &= \nabla_{\theta} \mathcal{L}_{FL} (\mathcal{I'}, f(\mathcal{X}, \theta)) \\
        &= \nabla_{\theta} ||\mathcal{I'} - f(\mathcal{X}, \theta)||^2_2 \\
        &= -2 \nabla_{\theta} f(\mathcal{X}, \theta) \cdot (\mathcal{I'} - f(\mathcal{X}, \theta))
    \end{aligned}
\end{equation}

Thus, we have:
\begin{equation} \label{eq:apd2}
        \nabla_{\mathcal{I'}} g(\mathcal{X}, \theta, \mathcal{I}') = -2 \nabla_{\theta}^T f(\mathcal{X}, \theta)
\end{equation}

and consequently:
\begin{equation} \label{eq:apd2b}
        \nabla^2_{\mathcal{I'}} g(\mathcal{X}, \theta, \mathcal{I}') = \mathbf{0}
\end{equation}

Therefore, by theorem \ref{thm:convexity}, there exists a global optimum for RAIFLE. Setting $\nabla_{\mathcal{I'}} \mathcal{L}_{atk} (\hat{\theta}, \theta')$ to $\mathbf{0}$ (from eq. \ref{eq:grad}), we have:
\begin{equation} \label{eq:apd3}
    \begin{aligned}
        & \nabla_{\mathcal{I'}} \mathcal{L}_{atk} (\hat{\theta}, \theta') = \mathbf{0} \\
        \iff & \nabla_{\mathcal{I'}} g(\mathcal{X}, \theta, \mathcal{I'}) \cdot (\hat{\theta} - g(\mathcal{X}, \theta, \mathcal{I'})) = \mathbf{0} \\
        \iff & \nabla_{\mathcal{I'}} g(\mathcal{X}, \theta, \mathcal{I'}) \cdot \hat{\theta} = \nabla_{\mathcal{I'}} g(\mathcal{X}, \theta, \mathcal{I'}) \cdot g(\mathcal{X}, \theta, \mathcal{I'}) \\
    \end{aligned}
\end{equation}

Substituting the results from eq. \ref{eq:apd1} and \ref{eq:apd2} into eq. \ref{eq:apd3}, we can now derive the closed-form solution for $\mathcal{I'}$:
\begin{equation} \label{eq:apd4}
    \begin{aligned}
        & \nabla_{\mathcal{I'}} \mathcal{L}_{atk} (\hat{\theta}, \theta') = \mathbf{0} \\
        \Leftrightarrow & \nabla_{\theta}^T f(\mathcal{X}, \theta) \cdot \hat{\theta} = 
        -2 \nabla_{\theta}^T f(\mathcal{X}, \theta) \nabla_{\theta} f(\mathcal{X}, \theta) \cdot (\mathcal{I'} - f(\mathcal{X}, \theta)) \\
        \Leftrightarrow & \mathcal{I'} = f(\mathcal{X}, \theta) -
        \frac{1}{2} \left(\nabla_{\theta}^T f(\mathcal{X}, \theta) \nabla_{\theta} f(\mathcal{X}, \theta) \right)^{-1} \nabla_{\theta}^T f(\mathcal{X}, \theta) \cdot \hat{\theta}
    \end{aligned}
\end{equation}

Note that $\nabla_{\theta}^T f(\mathcal{X}, \theta) \nabla_{\theta} f(\mathcal{X}, \theta)$ is invertible iff $\rank (\nabla_{\theta} f(\mathcal{X}, \theta)) = m$. Therefore, eq. \ref{eq:apd4} is a valid closed-form solution if and only if $\rank (\nabla_{\theta} f(\mathcal{X}, \theta)) = m$.

\subsection{Indirect Manipulation's Image Quality} \label{apd:img_quality}

To measure the similarity of the manipulated images to the originals in our federated OLTR experiment with image-based data (Section \ref{sec:eval_img_setup}), we calculated the peak signal-to-noise ratio (PSNR) and the structural similarity index measure (SSIM)~\cite{zhou2004ssim}.
Overall, the average SSIM ranges from 0.70 to 0.80 and the average PSNR ranges from 25 to 30 across all tested computer vision models (Table \ref{tab:img_similarity}).
These values indicate that the adversarial images are relatively similar to the originals but not without visible artifacts or distortions.
While our research is not focused on creating the most subtle manipulations, it is straightforward to include a regularization loss based on SSIM or PSNR in the manipulation process to improve image similarity, likely at the expense of attack performance.
We leave this for future research.

\begin{table}[h]
    \footnotesize
    \centering
    \caption{SSIM and PSNR of manipulated images and originals (rounded to 2nd decimal place)}
    \begin{tabular}{ccccc}
        \toprule
        Model & \makecell{SSIM\\Mean} & \makecell{SSIM\\Std.} & \makecell{PSNR\\Mean} & \makecell{PSNR\\Std.} \\
        \midrule
        ResNet18 & 0.70 & 0.06 & 25.80 & 1.39 \\
        RegNet Y 800MF & 0.80 & 0.06 & 29.36 & 1.27 \\
        DenseNet121 & 0.77 & 0.05 & 27.60 & 1.67 \\
        MNasNet 1.3 & 0.76 & 0.06 & 28.00 & 1.30 \\
        \bottomrule
    \end{tabular}
    \label{tab:img_similarity}
\end{table}

\subsection{Discussion of Additional Countermeasures} \label{apd:countermeasures}

We briefly discuss several hypothetical defense mechanisms against \name, including PIR, self-attacking, and alternative FL protocols. While these approaches are not yet as practical as the ones discussed in Section \ref{sec:countermeasures}, they are nonetheless theoretically interesting to consider.

\subsubsection{Private Information Retrieval} \label{manipulation:bypass_pir}

PIR~\cite{chor1998pir, kushilevitz1997pir} aims to allow a user to retrieve an item without the server knowing the identity of the item.
However, if a user participating in a PIR protocol also participates in IFL, then the privacy of their retrieved items can potentially be broken via our fingerprinting technique.
Consider a user $u$ and an item $t$: the server is interested in whether $u$ has retrieved $t$ or not.
Similar to bypassing SA, the server can fingerprint a feature of $t$ and zero out the values of that feature for all items other than $t$.
Thus, if the local update from user $u$ has a non-zero update to the feature parameters corresponding to the fingerprinted feature, it must be the case that $t$ was included in $u$'s training data.
Once again, this demonstrates the need to apply DP noise as the fingerprint would no longer be guaranteed to be preserved.

\subsubsection{Using \name as Guardrail}

Although we devise \name to demonstrate the increased privacy risks in IFL, it could also be repurposed as a form of ``sanity check'' before users send out their local updates.
Users would attack their own local updates and items to see if their true interactions can be reliably reconstructed or not.
If significant leakage is observed, users can choose not to share the local updates or to rerun their defense mechanisms with stronger privacy guarantees (e.g. reduce $\varepsilon$ in DP).
We envision self-attacking as a red-teaming technique to empirically assess privacy leakage in IFL.
This approach has two drawbacks: (1) users not being able to reconstruct their interactions does not necessarily mean the server would also be unable to, and (2) users will have to dedicate extra computation resources to run the attack.
To avoid a false sense of security, results should only be interpreted conservatively: success indicates likely leakage, while failure still leaves leakage potential.

\subsubsection{Alternative FL Protocols} \label{apd:fl_variations}

\name's threat model assumes a central FL server capable of controlling the interaction items.
One natural possible defense is to adopt a different FL architecture or protocol with no single centralized authority, targeting the core assumption of our attack.
For example, in the peer-to-peer gossip-based decentralized recommendation model~\cite{hegedus2019decentralized, li2024decentralized}, users exchange model weights/gradients and/or item interactions with other peers in the network, thus eliminating the need for a central server.
Hierarchical FL is another example where users are grouped into clusters and communicate exclusively with some statically or dynamically determined  ``leaders'' of the clusters~\cite{wang2021hierarchical, li2023hierarchical}.
While these scenarios can prevent \name thanks to their network topology, they often assume that peers are honest-but-curious and thus can still be vulnerable to Sybil attacks where malicious peers collude to gain an oversized influence on the FL network and uncover user interaction data~\cite{boenisch2023sybil}.
Additionally, even though the peers are also assumed to be able to communicate anonymously with one another, in reality, practical anonymous communication protocols often incur additional networking overhead (on top of the already increased communication expense from FL decentralization) and can still be subject to various deanonymization attacks~\cite{karunanayake2021tor}.
That said, we believe these ``serverless'' FL schemes hold great potential to further protect user privacy, and we hope to see more development and adoption from both academia and industry.

\newpage
\subsection{t-SNE Visualization of Gradients Under Manipulation}

\begin{figure}[h!]
    \centering
    \includegraphics[width=0.85\linewidth]{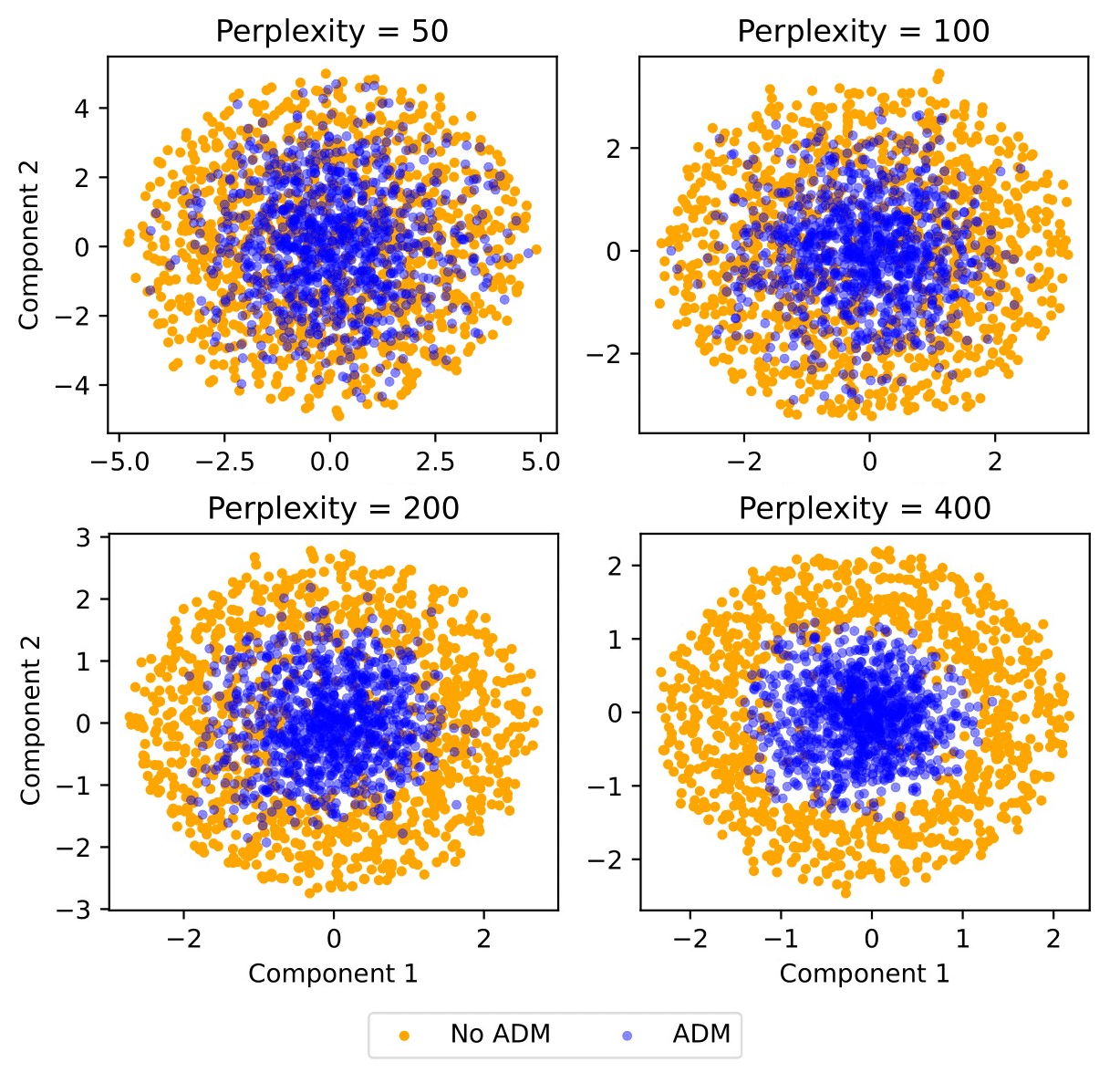}
    \caption{t-SNE visualization of the local FL gradients for neural FPDGD with 16 hidden units on MSLR-WEB10K, with and without ADM.}
    \label{fig:tsne_mslr10k}
\end{figure}

\begin{figure}[h!]
    \centering
    \includegraphics[width=0.85\linewidth]{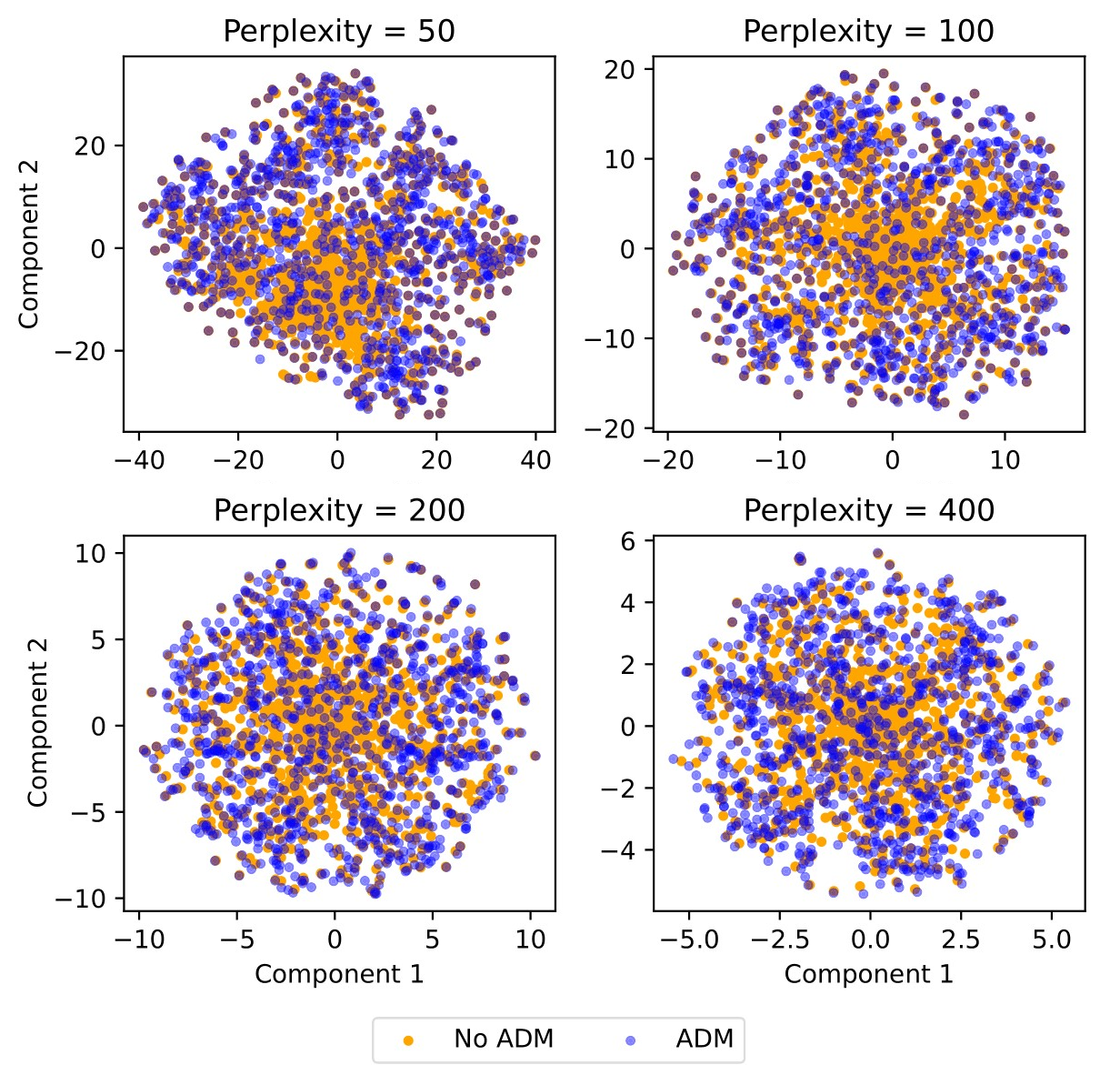}
    \caption{t-SNE visualization of the local FL gradients for image-based FOLTR with ResNet18 and a neural ranker (8 hidden units), with and without ADM.}
    \label{fig:tsne_resnet18}
\end{figure}

\newpage

\section{Artifact Appendix}

\subsection{Description \& Requirements}

\subsubsection{How to access}
\begin{itemize}
    \item DOI: \url{https://doi.org/10.17605/OSF.IO/HDSQR}
    \item Github: \url{https://github.com/dzungvpham/raifle}
\end{itemize}

\subsubsection{Hardware dependencies}
\begin{itemize}
    \item A commodity machine with at least 16GB of RAM and 30GB of storage.
    \item Optional but highly recommended: A CUDA-capable NVIDIA GPU with at least 8GB of VRAM (preferably 12-16GB)
\end{itemize}

\subsubsection{Software dependencies} 
\begin{itemize}
    \item A (x86-64) Unix-based OS. (Windows WSL will probably also work, but might need some installation modifications.)
    \item conda (such as Miniconda).
\end{itemize}

\subsubsection{Benchmarks}
We evaluate on the following datasets:
\begin{itemize}
    \item Recommendation: MovieLens-100K, STEAM-200K
    \item Learning-to-rank: MQ2008, MQ2007, MSLR-WEB10K
    \item Image: ImageNet 2012 (validation split)
\end{itemize}

\subsection{Artifact Installation \& Configuration}

We use conda to set up the environment.
Please see the README file in our repository for detailed instructions on how to set up as well as download the relevant datasets.

\subsection{Experiment Workflow}

Please refer to the Evaluation section.

\subsection{Major Claims}

\begin{itemize}
    \item (C1): \name achieves near-perfect AUC and significantly outperforms the state-of-the-art attack (IMIA) on federated recommendation (particularly the federated neural collaborative filtering algorithm), as demonstrated by experiment (E1) and reported in Tables \ref{tab:frs_results} and \ref{tab:raifle_vs_imia} of Section \ref{sec:eval_results_rec} for MovieLens-100K and STEAM-200K.
    \item (C2): \name outperforms traditional gradient inversion on federated online learning-to-rank, as demonstrated by experiments (E2) and (E3) and reported in Table \ref{tab:foltr_results} for MQ2007 and MSLR-WEB10K (Section \ref{sec:eval_results_foltr}) and Table \ref{tab:image_results} for ImageNet (Section \ref{sec:eval_results_image}).
\end{itemize}

\subsection{Evaluation}

All of our experiments are in IPython Jupyter Notebook inside the \code{code} folder.
We recommend using Visual Studio Code to run our experiments.
More detailed step-by-step instructions can be found in the README of our repository.
For artifact evaluation, we make some suggested modifications to our code (e.g., omit some experiment configurations and reduce the number of simulations/users) to keep the evaluation under the time limit.
Full-scale runs can be easily enabled by modifying our notebooks as instructed in the code comments.

Each experiment has the following structure:
\begin{itemize}
    \item The central FL server initializes a global FL model, performs manipulation on the FL training item features, then shares the model and items with users 
    \item Each user trains the global FL model locally using the manipulated data and their private interactions, applies differential privacy noise, then sends back the updated model weights to the server (directly or via secure aggregation).
    \item The server inverts the gradient updates to guess the private interactions.
\end{itemize}

\subsubsection{Experiment (E1)}
Federated Recommendation Systems: We run RAIFLE on the MovieLens-100K and STEAM-200K datasets with the Federated Neural Collaborative Filtering (FNCF) algorithm.

\textit{[Execution]}
Run all cells in \code{experiment\_rec.ipynb} in order. The default dataset is MovieLens-100K.
Cell 2 contains instructions on how to change the dataset. For artifact evaluation, we scaled down the number of users attacked to 30, which will take about ~30 minutes to finish.

\textit{[Results]} After the experiment completes, the raw results are saved as \code{output/rec\_metrics.csv} and a summary of the AUC and F1 score is printed out. The \code{name} column describes the configuration in the format: FNCF\_eps\_\{epsilon\}\_IMIA\_\{reg\_factor\},

where \{epsilon\} refers to the local DP $\varepsilon$ parameter (inf means no DP). IMIA\_0.0 means the IMIA defense is not applied, IMIA\_1.0 means the IMIA defense is applied with L1 regularization factor 1.0.

\begin{figure}[h!]
    \centering
    \includegraphics[width=1\linewidth]{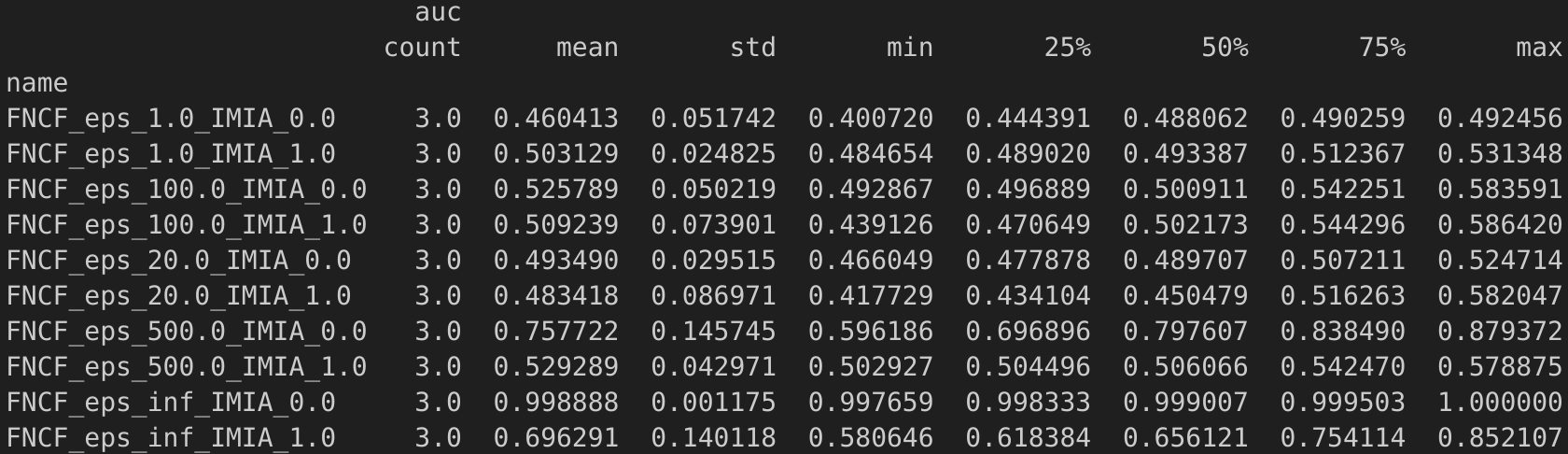}
    \caption{Example output of the federated RS experiment on MovieLens-100K for only 3 users}
    \label{fig:rec_metrics}
\end{figure}

\subsubsection{Experiment (E2)}
Federated Online Learning to Rank (FOLTR) (non-image): We run RAIFLE on the MQ2007, MQ2008, and MSLR-WEB10K datasets with the Federated Pairwise Differentiable Gradient Descent (FPDGD) algorithm.

\textit{[Execution]}
Run cell 1, 2, and 3 in \code{experiment\_ltr.ipynb}.
Cell 2 contains instructions on how to change the dataset and other configs. For artifact evaluation, the default configuration is a linear ranker + a neural net ranker with 16 hidden units, MQ2007 dataset, and 16 queries per user (this will take about 5 hours, a full run on MQ2007 can take more than 1 day, MSLR-WEB10K is much longer).

\textit{[Results]} After the experiment completes, the raw results are saved as \code{output/ltr\_metrics.csv} and a summary of the AUC is printed out. The \code{name} column describes the configuration in the format: \{model\_name\}\_\{click\_model\_name\}\_\{num\_query\}\_query \_eps\_\{epsilon\}\_\{key\},

where \{model\_name\} is `linear\_pdgd', `neural\_16\_pdgd', etc., \{click\_model\_name\} is either `informational' or `navigational', \{num\_query\} is the number of queries per user (e.g., 16), \{epsilon\} is the local DP epsilon (inf means no DP), and \{key\} is either 0.0 or 1.0, where 0.0 means no manipulation and 1.0 means full manipulation.

\begin{figure}[h!]
    \centering
    \includegraphics[width=1\linewidth]{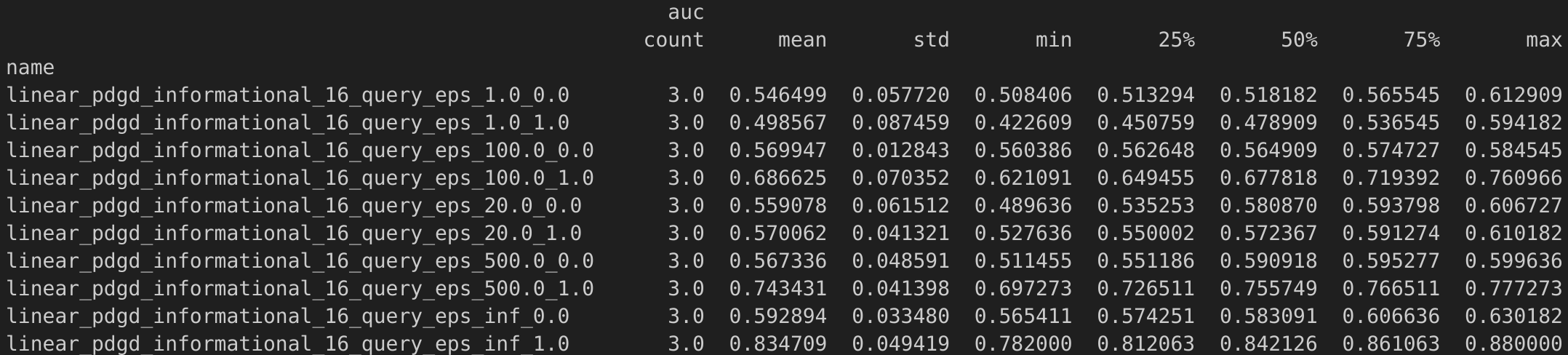}
    \caption{Example output of FOLTR experiment on MQ2007 for only 3 users with the default configuration.}
    \label{fig:ltr_metrics}
\end{figure}

\subsubsection{Experiment (E3)}
FOLTR with ImageNet: We run RAIFLE on the ImageNet dataset with a simple pointwise ranking algorithm. This is the only experiment where a GPU is recommended.

\textit{[Execution]}
Run cell 1, 2, 3, and 4 in \code{experiment\_ltr\_cv.ipynb}. The default configuration (scaled down for artifact evaluation) is ResNet18 as feature extractor, 30 rounds of simulation, and 5,000 images.
Cell 2 contains instructions on how to change the feature extractor.
Cell 3 generates the manipulated images. The batch size may need to be adjusted depending on how much GPU memory is available, e.g., 128 if 8GB, 256 if 12GB or more.
Cell 3 can take 2-3 hours with this configuration.
If a GPU is not available, we suggest using \code{raifle\_ltr\_cv\_colab.ipynb} with Google Colab's GPU runtime instead.
Cell 4 simulates the attack and can take 2-3 hours (without GPU).

\textit{[Results]} After the experiment completes, the raw results are saved as \code{output/ltr\_cv\_metrics.csv} and a summary of the AUC is printed out. The \code{name} column describes the configuration in the format: \{model\_name\}\_\{num\_items\}\_items\_eps\_\{epsilon\}\_\{key\}, where \{num\_items\} is 512, 1024, or 2048 (assuming ResNet18), \{epsilon\} is the local DP epsilon (inf means no DP), and \{key\} is `no\_adm' (no manipulation) or `adm\_opt' (RAIFLE).

\begin{figure}[h!]
    \centering
    \includegraphics[width=1\linewidth]{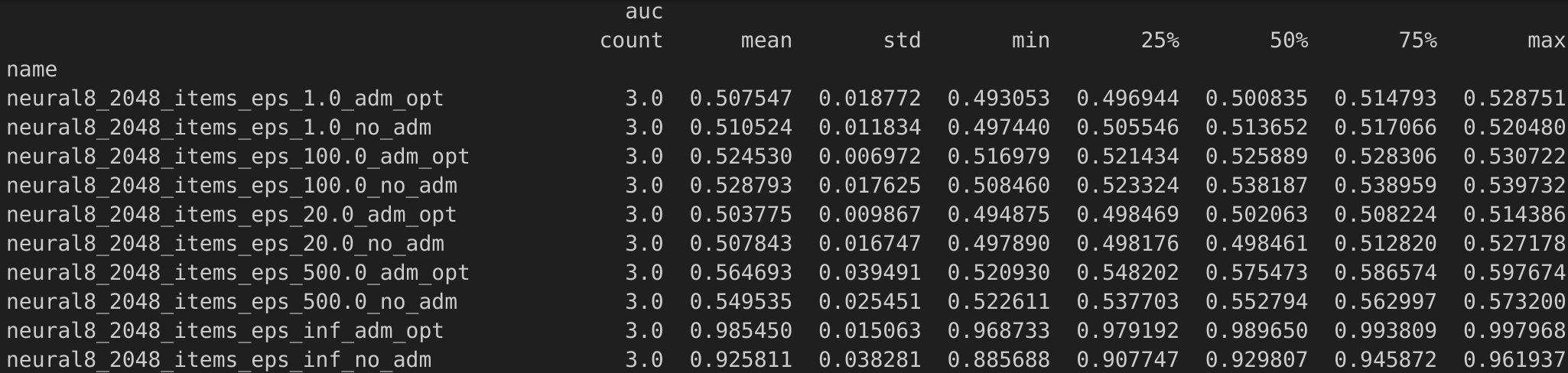}
    \caption{Example output of FOLTR experiment on ImageNet for only 3 simulation rounds with the default configuration.}
    \label{fig:ltr_cv_metrics}
\end{figure}